\PassOptionsToPackage{rgb}{xcolor}
\documentclass[a4paper,USenglish,cleveref,autoref,thm-restate,pdfa]{lipics-v2021}

\hideLIPIcs

\nolinenumbers 

\newcommand{\system}{\textsc{TypeWeaver}}
\newcommand{\ie}{i.e.}
\newcommand{\eg}{e.g.}
\newcommand{\deeptyper}{Deep\-Typer}
\newcommand{\lambdanet}{Lambda\-Net}
\newcommand{\incoder}{In\-Coder}
\newcommand{\es}{ECMA\-Script}

\lstdefinelanguage{JavaScript}[]{Java}{
morekeywords={await,debugger,delete,export,import,false,function,in,let,null,true,typeof,var,with,yield},
deletekeywords={assert,strictfp}
}

\newcommand{\totalpkgs}{513}

\title{Do Machine Learning Models Produce TypeScript Types That Type Check?}

\author{Ming-Ho Yee}{Northeastern University, Boston, MA, USA}{mh@mhyee.com}{https://orcid.org/0000-0002-8008-8481}{}
\author{Arjun Guha}{Northeastern University, Boston, MA, USA \and Roblox Research, San Mateo, CA, USA}{a.guha@northeastern.edu}{https://orcid.org/0000-0002-7493-3271}{}

\authorrunning{M.-H. Yee and A. Guha}
\Copyright{Ming-Ho Yee and Arjun Guha}



\supplementdetails[linktext={\\https://doi.org/10.4230/DARTS.9.2.5}, subcategory={ECOOP 2023 Artifact Evaluation approved artifact}]{Software}{https://doi.org/10.4230/DARTS.9.2.5}
\supplementdetails[subcategory={Artifact Evaluation approved artifact}]{Software}{https://doi.org/10.5281/zenodo.7662708}
\supplementdetails[swhid={swh:1:dir:34399ede560aa59cfe736bf9994185d54b8c2e7e;origin=https://github.com/nuprl/TypeWeaver;visit=swh:1:snp:4a7a6a8efedaefb20b8678af76feceed8a25eb25;anchor=swh:1:rev:a0edcd84e044f3f5d4ee76e648c0f09b3dc2b71e},subcategory={code repository}]{Software}{https://github.com/nuprl/TypeWeaver}

\acknowledgements{We thank Northeastern Research Computing and the New England Research Cloud for providing computing resources; and Leif Andersen, Luna Phipps-Costin, Donald Pinckney, and the anonymous reviewers for their feedback.}

\funding{This work is partially supported by the National Science Foundation grant CCF-2102291.}

\EventEditors{Karim Ali and Guido Salvaneschi}
\EventNoEds{2}
\EventLongTitle{37th European Conference on Object-Oriented Programming (ECOOP 2023)}
\EventShortTitle{ECOOP 2023}
\EventAcronym{ECOOP}
\EventYear{2023}
\EventDate{July 17--21, 2023}
\EventLocation{Seattle, Washington, United States}
\def\EventLogoHeight{41}
\EventLogo{badge-available-reusable}
\EventLogoURL{https://doi.org/10.4230/DARTS.9.2.5}
\SeriesVolume{263}
\ArticleNo{37}

\begin{document}

\maketitle
\vspace{0.5\baselineskip}
\enlargethispage{-0.5\baselineskip}
\begin{abstract}
Type migration is the process of adding types to untyped code to gain assurance
at compile time. TypeScript and other gradual type systems facilitate type
migration by allowing programmers to start with imprecise types and gradually
strengthen them. However, adding types is a manual effort and several migrations
on large, industry codebases have been reported to have taken several years. In the
research community, there has been significant interest in using machine learning
to automate TypeScript type migration. Existing machine learning models report
a high degree of accuracy in predicting individual TypeScript type annotations.
However, in this paper we argue that accuracy can be misleading, and
we should address a different question: can an automatic type migration tool
produce code that passes the TypeScript type checker?

We present \textsc{TypeWeaver}, a TypeScript type migration tool  that can be
used with an arbitrary type prediction model. We evaluate \textsc{TypeWeaver}
with three models from the literature: DeepTyper, a recurrent neural network;
LambdaNet, a graph neural network; and InCoder, a general-purpose,
multi-language transformer that supports fill-in-the-middle tasks. Our tool
automates several steps that are necessary for using a type prediction model,
including (1)~importing types for a project's dependencies; (2)~migrating
JavaScript modules to TypeScript notation; (3)~inserting predicted type
annotations into the program to produce TypeScript when needed; and
(4)~rejecting non-type predictions when needed.

We evaluate \textsc{TypeWeaver} on a dataset of 513 JavaScript packages,
including packages that have never been typed before. With the best type
prediction model, we find that only 21\% of packages type check,
but more encouragingly, 69\% of files type check successfully.
\end{abstract}

\section{Introduction}\label{sec:introduction}

Gradual typing allows programmers to freely mix statically and dynamically
typed code. This makes it possible to add static types to a large program
incrementally, and slowly reap the benefits of static typing without requiring a
complete rewrite of an existing codebase at
once~\cite{guha:flowtypes,siek:gtlc,th:typed-scheme}. Over the past decade,
gradual typing has proliferated, and there are now gradually typed versions of
several mainstream
languages~\cite{bierman:ts,bonnaire-sergeant:typed-clojure,cassola:gradual-elixir,chauduri:fb-flow,lu:static-python,pyre,ottoni:hhvm,th:typed-scheme,stripe:sorbet}.

TypeScript is a widely used gradually typed language, and a syntactic
superset of JavaScript. Programmers can write their code in TypeScript, benefit
from static typing, and then compile to JavaScript.
However, the process of \emph{migrating} an untyped JavaScript program to
TypeScript has remained a labor-intensive manual effort in practice.
For example, Airbnb engineers took more than two years to add TypeScript type
annotations to 6 million lines of JavaScript~\cite{airbnb:ts-migrate}, and
there are several other accounts of multi-year TypeScript migration efforts~\cite{heap:ts,netflix:ts,abacus:ts,quip:ts,slack:ts}.

To address this problem, there has been significant research interest in
using machine learning to predict TypeScript types. Machine
learning seems attractive because TypeScript has language
features (\eg{}, \texttt{eval}) that are very difficult to accommodate with
traditional, constraint-based approaches. Moreover, there is a significant
quantity of open-source TypeScript that is available to serve as training
data for a machine learning model. Over the last few years, advances in
model architectures and high-quality training data have led to type annotation prediction
with high accuracy on individual type
annotations~\cite{hellendoorn:dlti,jesse:diversetyper,jesse:typebert,pandi:opttyper,wei:lambdanet}.

However, in this paper we argue that accuracy can be misleading, and that
predicting individual type annotations is just the first step of migrating a codebase from
JavaScript to TypeScript.
We should address a different question: \emph{can an automatic type migration
tool produce code that type checks?}
If so, we prefer type annotations that are non-trivial and
useful (\ie{}, annotations that are not just \texttt{any}).
On the other hand, if the code does not type check, it may have too many errors, which can overwhelm a user who may
just turn off the tool.
Moreover, it may not be feasible to fix the type errors automatically, since
type errors refer to code locations whose typed terms are used, and not
necessarily to faulty annotations.

To answer the type checking question, we present \system{}, a TypeScript type
migration tool that can be used with an arbitrary type prediction model.
Our evaluation employs three models from the literature:
\deeptyper{}~\cite{hellendoorn:dlti}, a recurrent neural network;
\lambdanet{}~\cite{wei:lambdanet}, a graph neural network; and
\incoder{}~\cite{fried:incoder}, a general-purpose, multi-language
transformer that supports fill-in-the-middle tasks. Our tool automates several
steps that are necessary for using a type prediction model, including:
\begin{description}
\item [Importing type dependencies]
Before migrating a JavaScript project, we must ensure that its dependencies
are typed. This means transitively migrating dependencies, or ensuring that the
dependencies have TypeScript interface declaration~(\texttt{.d.ts}) files available.

\item [Module conversion]
JavaScript code written for Node.js may use either the CommonJS or \es{} module
system.
However, when migrated to TypeScript, only \es{} modules preserve type information.
Thus, to fully benefit from static type checking, code written with CommonJS
modules should be refactored to use \es{} modules.

\item [Type weaving]
Models that assign type labels to variables do not update the JavaScript source
to include type annotations. Therefore, to ask if a program type
checks, we must ``weave'' the predicted type annotations with the original
JavaScript source to produce TypeScript.

\enlargethispage{\baselineskip}
\item[Rejecting non-type predictions]
Models that predict type annotations as in-filled sequences
of tokens can easily produce token sequences that are not syntactic types.
These predictions need to be rejected or cleaned for type prediction to work.
\end{description}
After completing these tasks, it is then possible to type check the resulting
TypeScript program and evaluate the effectiveness of \system{}.

\bigskip

Our contributions are the following:
\begin{itemize}

\item We describe \system{}, a TypeScript type migration tool for evaluating type
prediction models, which automates several prerequisite
steps~(\cref{sec:approach}).

\item We provide a dataset of \totalpkgs{} JavaScript packages, which are a
subset of the top 1,000 most downloaded packages from the npm Registry.
Our dataset includes packages without known type annotations,
\ie, code that has never been used before to evaluate type prediction
models~(\cref{sec:eval-dataset}).

\item We report the success rate of type checking. We answer the questions of
how many packages type check, how many files type check, how many type
annotations are trivial, and whether the predicted types match human-written
types~(\cref{sec:eval-type-checking}).

\item We discuss the common kinds of errors that result from a type
migration~(\cref{sec:eval-error-analysis}).

\item We compare the results of a type migration before and after converting
to the \es{} module system~(\cref{sec:eval-cjs-vs-es}).

\item Finally, we examine four packages as case studies, to showcase other
difficulties that arise during type migration~(\cref{sec:case-study}).

\end{itemize}

\section{Background}\label{sec:background}

In this section, we first provide background on the type migration problem and
contrast it to type inference. We then discuss deep-learning-based type
annotation prediction, focusing on the tools that we have used with \system.

\subsection{Type Migration vs.\ Type Inference}\label{sec:bg-type-migration}

\begin{figure}[t]
\centering
\begin{subfigure}[t]{0.45\textwidth}
\begin{lstlisting}
function f(x) { return x+x; }#\label{line:js-param}#
f(1)    // returns 2#\label{line:js-num}#
f("a")  // returns "aa"#\label{line:js-str}#
    \end{lstlisting}
\caption{JavaScript function that adds or concatenates its argument to
itself.}\label{fig:js-plus-operator}
\end{subfigure}
\qquad
\begin{subfigure}[t]{0.45\textwidth}
\begin{lstlisting}
var point = {};#\label{line:js-point}#
point.x = 42;#\label{line:js-point-x}#
point.y = 54;#\label{line:js-point-y}#
    \end{lstlisting}
\caption{JavaScript that creates a ``point'' object.}\label{fig:js-point}
\end{subfigure}
\caption{JavaScript code that cannot be easily typed in TypeScript.}\label{fig:js-example}
\end{figure}

Type inference is related to, but distinct from, type migration.
The goal of type inference is to \emph{reconstruct} the types of variables, expressions, and functions, where some or all the type annotations are missing.
In other words, the language is statically typed and the types exist implicitly within the program, so the type inference algorithm computes the missing annotations.
Furthermore, inference can frequently compute the \emph{principal type} of a variable, expression, or function, \ie, the most general type.
As a result, there is a single answer for the type of a variable, expression, or function.
Additionally, in languages that support type inference, the inferred type
annotations are well defined and not added to the program text.

In this paper, we use \emph{type migration} to describe the problem of migrating
a program from an untyped language to a typed language, \eg, from JavaScript to
TypeScript, a process that may require refactoring in addition to type
inference.
The type migration process starts from an untyped program without type annotations:
there is no type information available, beyond the basic information available
from literal values, operators, and control-flow statements, so type definitions may need to be inserted into the program.
Furthermore, multiple type annotations may be valid, rather than having a single principal type.
For example, \cref{fig:js-plus-operator} shows a JavaScript function where the parameter \texttt{x} on \cref{line:js-param} could be annotated as \texttt{number} or \texttt{string}; without additional context, both annotations are valid.
The example illustrates another challenge of type migration: \texttt{f} is called on \cref{line:js-num} with a number and \cref{line:js-str} with a string, so the only valid type annotation for \texttt{x} is \texttt{any}.\footnote{The union type \texttt{number | string} produces a type error in the function.}
This satisfies the TypeScript compiler's type checker, but may not be a helpful annotation in terms of documentation.

The type migration problem for TypeScript has several additional challenges and we highlight some of them here.
TypeScript has a structural type system, which makes it even harder to determine the right annotation.
Furthermore, structural types are often verbose, making it difficult for a programmer to understand the code, which defeats one of the benefits of a static type system.
Another difficulty is that JavaScript code can be too dynamic to fit within
TypeScript's type system, \eg, there is no good way to handle \texttt{eval}, other than using the \texttt{any} annotation as an escape hatch.
Finally, certain idioms and patterns in JavaScript code do not fit TypeScript and need to be refactored.
For instance, consider \cref{fig:js-point}, which initializes a ``point'' object in JavaScript.
\cref{line:js-point} initializes \texttt{point} to an empty object, and \cref{line:js-point-x,line:js-point-y} set the \texttt{x} and \texttt{y} properties.
However, this cannot be easily typed in TypeScript,\footnote{The correct, but awkward, type annotation is \texttt{\{x?: number, y?: number\}}, which declares \texttt{x} and \texttt{y} as optional properties. If \texttt{x} or \texttt{y} were required, the assignment on \cref{line:js-point} would be a type error. Alternatively, \texttt{any} is valid, but unhelpful.} and it is more appropriate to rewrite the code to use TypeScript classes.

In this landscape of challenges, recent work has focused on the narrower problem
of assigning type annotations to TypeScript code, in particular, using deep learning approaches.
We examine some of these approaches in the next subsection.

\subsection{Deep-Learning-Based Type Annotation Prediction}\label{sec:bg-predicting-types}

We focus on JavaScript and TypeScript, since there have been a variety
of proposed type prediction models for those languages. We evaluate three of
them here: \deeptyper{}~\cite{hellendoorn:dlti}, \lambdanet{}~\cite{wei:lambdanet},
and \incoder{}~\cite{fried:incoder}.

\deeptyper{} was the first deep neural network for TypeScript type prediction,
and uses a \emph{bidirectional recurrent neural network} architecture.
\lambdanet{} was another early approach, and it uses a \emph{graph neural
network} architecture. \incoder{} is a recent \emph{large language model} that
predicts arbitrary code completions, and while not trained specifically to
predict type annotations, its ``fill in the middle'' capability makes it ideal
for that task. All three models use training data from public code repositories.

We require a system that takes a JavaScript project as input and outputs a
type-annotated TypeScript project. \deeptyper{} and \lambdanet{} output a probability
distribution of types for each identifier, which we must then ``weave'' into the
original JavaScript source to produce TypeScript; we describe this technique in
\cref{sec:approach-type-weaving}. \incoder{} is a general-purpose,
multi-language transformer, so we implemented a front end to use \incoder{} to
predict type annotations and output TypeScript. Currently, our front end only
supports type predictions for function parameters; we describe our
implementation in \cref{sec:approach-inference-ic}.

Our approach can be adapted to work with any type prediction model. Older models
may require some work to adapt their outputs, but our \incoder{} front end can
easily be extended to support other fill-in-the-middle models, such as OpenAI's
model~\cite{bavarian:openai} and SantaCoder~\cite{benallal:santacoder}.

\subsubsection{DeepTyper}\label{sec:bg-deeptyper}

\deeptyper{}~\cite{hellendoorn:dlti} predicts types for variables, function
parameters, and function results using a fixed vocabulary of types,
\ie{}, it cannot predict types declared by the program under
analysis unless those types were observed during training.
\deeptyper{} treats type inference as a machine translation problem from one language (unannotated TypeScript) to another (annotated TypeScript).
Specifically, it uses a model based on a \emph{bidirectional recurrent neural network} architecture to translate a sequence of tokens into a sequence of types: for each identifier in the source program, \deeptyper{} returns a probability distribution of predicted types.
Because the input token sequence is perfectly aligned with the output type
sequence, this task can also be considered a sequence annotation task, where an
output type is expected for every input token.\footnote{The \deeptyper{}
architecture must classify \emph{every} input token, including ones
where an output type does not make sense, such as \texttt{if}, \texttt{(},
\texttt{)}, and even whitespace. \deeptyper{} filters out these predictions, so
a user will never observe these meaningless types.}
However, this approach treats each input token as independent from the
others, \ie{},
a source variable may be referenced multiple times and each occurrence may have
a different type.
To mitigate this, \deeptyper{} adds a consistency layer to the neural network, which encourages -- but cannot enforce -- the model to treat multiple occurrences of the same identifier as related.

\deeptyper{}'s dataset is based on the top 1,000 most starred TypeScript projects on GitHub, as of February 2018.
After cleaning to remove large files (those with more than 5,000 tokens) and projects that contained only TypeScript declaration files, the dataset was left with 776 TypeScript projects (containing about 62,000 files and about 24 million tokens), which were randomly split into 80\% (620 projects) training data, 10\% (78 projects) validation data, and 10\% (78 projects) test data.
Further processing and cleaning of rare tokens resulted in a final vocabulary of 40,195 source tokens and 11,830 types.

The final training dataset contains both identifiers and types, where each identifier has an associated type annotation; this includes annotations inferred by the TypeScript compiler that were not manually annotated by a programmer.
The testing dataset contains type annotations and no identifiers; specifically, the type annotations added by programmers are associated with their declaration sites, and all other sites are associated with ``no-type.''
As a result, \deeptyper{}'s predictions are evaluated against the handwritten type annotations, rather than all types in a project.

\subsubsection{LambdaNet}\label{sec:bg-lambdanet}

Like \deeptyper{}, \lambdanet{}~\cite{wei:lambdanet} predicts type annotations for variables, function parameters, and function returns: it takes an unannotated TypeScript program and outputs a probability distribution of predicted types for each declaration site.
\lambdanet{} improves upon two limitations of \deeptyper{}.
First, it predicts from an open vocabulary, beyond the types that were observed during training; \ie{}, it can predict user-defined types from within a project.
Second, it only produces type predictions at declaration sites, rather than at every variable occurrence; in other words, multiple uses of the same variable will have a consistent type.

\lambdanet{} uses a \emph{graph neural network} architecture and represents a source program as a so-called \emph{type dependency graph}, which is computed from an intermediate representation of TypeScript that names each subexpression.
The type dependency graph is a hypergraph that encodes program type variables as nodes, and relationships between those variables as labeled edges.
By encoding type variables, \lambdanet{} makes a single prediction over all occurrences of a variable, rather than a prediction for each instance of a variable.
Furthermore, the edges encode logical constraints and contextual hints.
Logical constraints include subtyping and assignment relations, functions and
calls, objects, and field accesses, while contextual hints include variable
names and usages.
Finally, \lambdanet{} uses a \emph{pointer network} to predict type annotations.

\lambdanet{}'s dataset takes a similar approach to \deeptyper{}: they selected
the 300 most popular TypeScript projects from GitHub that contained 500--10,000
lines of code, and had at least 10\% of type annotations that referred to user-defined types.
The dataset has about 1.2 million lines of code, and only 2.7\% of the code is duplicated.
The 300 projects were split into three sets: 200 (67\%) for training data, 40 (13\%) for validation data, and 60 (20\%) for test data.
The vocabulary was split into library types, which consist of the top 100 most
common types in the training set, and user-defined types, which are all the
classes and interfaces defined in source projects.
Similar to \deeptyper{}, \lambdanet{}'s predictions are evaluated against the handwritten annotations that were added by programmers.

\subsubsection{InCoder}\label{sec:bg-incoder}

\incoder{}~\cite{fried:incoder} is a \emph{large language model}~(LLM) for
generating arbitrary code that is trained with a \emph{fill-in-the-middle}~(FIM) objective on a corpus of several programming languages, including
TypeScript.

\incoder{}'s corpus consists of permissively licensed, open-source
code from GitHub and GitLab, as well as Q\&A and comments from Stack Overflow.
This raw data is filtered to exclude: (1)~code that is duplicated; (2)~code
that is not written in one of 28 languages; (3)~files that that
are extremely large or contain very few alphanumeric characters; (4)~code
that is likely to be compiler generated; and (5)~certain code
generation benchmarks. The result is about 159 GB of code, which is dominated
by Python and JavaScript. TypeScript is approximately 4.5 GB of the training
data.

We describe \incoder{} in more depth in \cref{sec:approach-inference-ic},
where we present what is necessary to use it as a type annotation prediction
tool for TypeScript.

\subsubsection{Evaluating on Accuracy}\label{sec:bg-accuracy}

The main evaluation criteria for the type annotation prediction task is accuracy: what is the likelihood that a predicted type annotation is correct?
Correct means the prediction \emph{exactly} matches the ground truth, which is the handwritten type annotation at that location.
Accuracy is typically measured as top-$k$ accuracy, where a prediction is deemed correct if any of the top~$k$ most probable predictions is correct.
For our evaluation, we would like a result that ``just works,'' \ie{}, a program that type checks.
Therefore, we are only interested in top-1 accuracy, since we take the top guess as the only prediction.

\deeptyper{}'s test dataset makes up 10\% (78 projects) of its original corpus and contains only the annotations that were manually added by programmers.
Predictions are compared against this ground truth dataset, and \deeptyper{} reports a top-1 accuracy of 56.9\%.
Because \deeptyper{} may predict different types for multiple occurrences of the same variable, the authors also report an inconsistency metric: 15.4\% of variables had multiple type predictions.

\lambdanet{} also compares its predictions against a ground truth of handwritten type annotations, but they use a different corpus and split 20\% (60 projects) for the test dataset.
\lambdanet{} can predict user-defined types, so the evaluation reports two sets of results: a top-1 accuracy of 75.6\% when predicting only common library types, and a top-1 accuracy of 64.2\% when predicting both library and user-defined types.

\incoder{} was not designed specifically to predict TypeScript type annotations, but the authors report an experiment to predict only the result types for Python functions.
For this task, \incoder{} was evaluated on a test dataset of 469 functions, which was constructed from the CodeXGLUE dataset; \incoder{} achieved an accuracy of 58.1\%.

\medskip

However, we argue that accuracy is not the right metric for evaluating a type
prediction model. As a first step, we would like to type check the TypeScript
project. Additionally, when migrating a JavaScript project to
TypeScript, there is frequently no ground truth of handwritten type annotations;
instead, the ground truth is what the compiler accepts. This condition is much
stronger than accuracy, as even a single, incorrect type annotation causes a
package to fail to type check. On the other hand, less precise type annotations
(\eg{}, \texttt{any}) and equivalent annotations (\eg{},
\texttt{number | string} vs.\ \texttt{string | number}) may be accepted,
despite not matching the ground truth exactly.

In the next section, we describe our approach for type migration and evaluating type prediction models.

\section{Approach}\label{sec:approach}

We take an end-to-end approach to type migration, starting from an untyped
JavaScript project and finishing with a type-annotated TypeScript project that we try to type check.
The first step, which is optional, is to convert from CommonJS modules to \es{} modules.
Next, we invoke one of the type prediction models: \deeptyper{} and \lambdanet{}
produce type predictions, while \incoder{}, with a front end we implemented, produces TypeScript.
Because \deeptyper{} and \lambdanet{} do not produce TypeScript, we perform a
step that we call \emph{type weaving}, which combines type predictions with
the original JavaScript source and outputs TypeScript.
Finally, we run the TypeScript compiler to type check the now migrated TypeScript project.

\subsection{CommonJS to \es{} Module Conversion}\label{sec:approach-cjs-to-es}

The first step is to convert projects from CommonJS module notation to \es{} module notation.
This step is not necessary for type prediction, but is important for the type checking evaluation, as only
\es{} modules preserve type information across module boundaries.
Because we treat this step as optional, our dataset has two versions:
the original projects, which may use CommonJS or \es{} modules, and a final version that only uses \es{} modules.

\begin{figure}[t]
\centering
\begin{subfigure}[t]{0.45\textwidth}
\begin{lstlisting}
// a.js
var x = 2;    // private#\label{line:cjs-a-x}#

module.exports.foo = 42;#\label{line:cjs-a-foo}#
module.exports.f = (i) => i+x;#\label{line:cjs-a-f}#
    \end{lstlisting}
\caption{CommonJS: \texttt{a.js} exports \texttt{foo} and \texttt{f}.}\label{subfig:cjs-a}
\end{subfigure}
\qquad
\begin{subfigure}[t]{0.45\textwidth}
\begin{lstlisting}
// b.js
var a = require('./a.js');#\label{line:cjs-b-require}#

console.log(a.foo);  // 42#\label{line:cjs-b-foo}#
console.log(a.f(1)); // 3#\label{line:cjs-b-f}#
    \end{lstlisting}
\caption{CommonJS: \texttt{b.js} imports the module \texttt{a.js}.}\label{subfig:cjs-b}
\end{subfigure}

\bigskip

\begin{subfigure}[t]{0.45\textwidth}
\begin{lstlisting}
// a.mjs
var x = 2;    // private

export var foo = 42;
export var f = (i) => i+x;
    \end{lstlisting}
\caption{\es{}: \texttt{a.mjs} exports \texttt{foo} and \texttt{f}.}\label{subfig:esm-a}
\end{subfigure}
\qquad
\begin{subfigure}[t]{0.45\textwidth}
\begin{lstlisting}
// b.mjs
import {foo,f} from './a.mjs';

console.log(foo);  // 42
console.log(f(1)); // 3
    \end{lstlisting}
\caption{\es{}: \texttt{b.mjs} imports \texttt{foo} and \texttt{f}.}\label{subfig:esm-b}
\end{subfigure}
\caption{An example comparing the CommonJS and \es{} module systems.}\label{fig:cjs-es-example}
\end{figure}

Most Node.js packages use the CommonJS module system, which was the original module system for Node.js and remains the default.
\cref{subfig:cjs-a,subfig:cjs-b} show an example of the CommonJS module system,
where files \texttt{a.js} and \texttt{b.js} implement separate modules.
In this example, \texttt{a.js} sets the \texttt{foo} and \texttt{f} properties of the special \texttt{module.exports} object.
Local variables like \texttt{x} are private and not exported.
On \cref{line:cjs-b-require}, \texttt{b.js} uses the Node.js function \texttt{require} to load module \texttt{a.js} into the local variable \texttt{a}.
As a result, \texttt{a} takes on the value of the \texttt{module.exports} object set by \texttt{a.js}, and both \texttt{foo} and \texttt{f} are available as properties of \texttt{a}.

\es{} 6 introduced a new module system, referred to as \es{} modules.
Node.js supports \es{} modules when using the \texttt{.mjs} extension or setting a project-wide configuration in the \texttt{package.json} file.
\cref{subfig:esm-a,subfig:esm-b} show the same program as before, but rewritten to use \es{} modules.
In this example, \texttt{a.mjs} directly exports \texttt{foo} and \texttt{f}, rather than writing to a special \texttt{module.exports} object.
Then, \texttt{b.mjs} directly imports the names with the \texttt{import} statement instead of loading an object.

TypeScript supports both CommonJS and \es{} modules, depending on the project configuration.
However, CommonJS modules in TypeScript are untyped; specifically, \texttt{require} is typed as a function that returns \texttt{any}.
Therefore, even if a module has type annotations for the variables and functions it exports, those annotations are lost when the module is imported.
On the other hand, with \es{} modules, the \texttt{import} statement preserves the type annotations of names it imports.

In order to make use of the most type information available, we would like to use \es{} modules in our evaluation.
Therefore, we use the
\texttt{cjs-to-es6} tool\footnote{\url{https://github.com/nolanlawson/cjs-to-es6}}
to transform our dataset to use \es{} modules.
The conversion tool is not perfect, and in particular has difficulty when
\texttt{require} is used to dynamically load a module. Some of these cases could
be fixed manually, but many are genuine uses of dynamic loading in JavaScript.

\subsection{Type Annotation Prediction}\label{sec:approach-type-inference}

The next step is to invoke a deep learning model to predict type annotations for a JavaScript project.
\deeptyper{} and \lambdanet{} require an additional step, which we call
\emph{type weaving}, to produce TypeScript, while \incoder{}, with our front end, outputs TypeScript directly.

We use the pretrained \deeptyper{} model available from its GitHub
repository,\footnote{\url{https://github.com/DeepTyper/DeepTyper/tree/master/pretrained}}
which is not identical to the model used in the \deeptyper{} paper.
\deeptyper{} reads in a JavaScript file, and for each identifier, predicts the
top five most likely types, outputting the result in comma-separated values~(CSV)
format.

We use the pretrained \lambdanet{} model available from its GitHub
repository,\footnote{\url{https://github.com/MrVPlusOne/LambdaNet/tree/ICLR20}}
specifically the model that supports user-defined types.
\lambdanet{} reads in a directory containing a JavaScript project, and predicts
the top five most likely types for each variable and function declaration.
We modify \lambdanet{} to output in CSV format.

\deeptyper{} predicts types for all identifiers in the program, including program locations that do not allow type annotations.
Therefore, type weaving must also ensure that type annotations are applied
correctly, \ie, only to variable declarations, function parameters, and function results.
\lambdanet{} predicts types for variable and function declarations, and in the
correct locations; however, type weaving is still required to produce TypeScript
code.
Our \incoder{} front end does not require type weaving, but only supports type
predictions for function parameters.
We use the pretrained \incoder{} model available from
Hugging Face.\footnote{\url{https://huggingface.co/facebook/incoder-6B}}

\subsubsection{InCoder}\label{sec:approach-inference-ic}

\begingroup
\setlength\fboxsep{0pt}

\begin{figure}[t]
\centering
\begin{subfigure}[t]{0.45\textwidth}
\begin{lstlisting}
function f(x) {
  return x + 1;
}
\end{lstlisting}
\caption{An example program.}
\end{subfigure}
\qquad
\begin{subfigure}[t]{0.45\textwidth}

\begin{lstlisting}
function f(x: #\colorbox{yellow}{<|mask:0|>}#) {
  return x + 1;
}#\colorbox{yellow}{<|mask:1|><|mask:0|>}
\end{lstlisting}
\caption{Preparing code for generation.}
\end{subfigure}

\bigskip

\begin{subfigure}[t]{0.45\textwidth}

\begin{lstlisting}
function f(x: <|mask:0|>) {
  return x + 1;
}<|mask:1|><|mask:0|>
#\colorbox{yellow}{number, y: number<|endofmask|>}#
\end{lstlisting}
\caption{\incoder{} often produces extra tokens after the type. Here it produces
a new parameter that is not in the original program.}
\label{incoder-hallucination}
\end{subfigure}
\qquad
\begin{subfigure}[t]{0.45\textwidth}
\begin{lstlisting}
function f(x: number) {
  return x + 1;
}
\end{lstlisting}
\caption{We select a prefix of the generated program that is a syntactically valid
TypeScript type.}
\end{subfigure}
\caption{Generating types with \incoder{}.}\label{fig:incoder-infill}
\end{figure}

\endgroup

\incoder{} is trained to generate code in the middle of a program, conditioned
on the surrounding code. To train on a single example (a file of code), the
training procedure replaces a randomly selected contiguous span of tokens with a
\emph{mask sentinel} token. It appends the mask sentinel to the end of the
example, followed by the tokens that were replaced and a special \emph{end-of-mask}
token. The model is then trained as a left-to-right language model.
This approach generalizes to support several, non-overlapping masked spans, and
its training examples have up to 256 randomly selected masked spans, though
the majority have just a single masked span.

In principle one could give \incoder{} a program with up to 256 types
to generate at once. However, we found that \incoder{} is more successful
generating a single type at a time, and with a limited amount of context.
To generate a type annotation with \incoder{}  we (1)~insert the mask sentinel
token at the insertion point; (2)~add the mask sentinel to the end of the file;
(3)~generate at the end of the file until the model produces the end-of-mask
token; (4)~move the generated text to the insertion point; and (5)~remove all
sentinels.

\Cref{fig:incoder-infill} shows an example of generating a type annotation.
However, a problem that we frequently encountered is that \incoder{}
sometimes generates more than just a single type. For instance,
\cref{incoder-hallucination} shows an example where \incoder{} generates a
new parameter that is not in the original program. The simplest approach is
to reject this result and get \incoder{} to re-generate completions until
it produces a type. However, we found that it is far more efficient to
accept a prefix of the generated code if it is a syntactically valid type,
which we check with a TypeScript parser in the generation loop.

\subsection{Type Weaving}\label{sec:approach-type-weaving}

To produce type-annotated TypeScript code, we use a process we call \emph{type weaving} to combine type predictions with the original JavaScript code.
Type weaving takes two files as input: a JavaScript source file and an associated CSV file with type predictions.
The type weaving program parses the JavaScript source into an abstract syntax
tree~(AST), and then traverses the AST and CSV files simultaneously, using the
TypeScript compiler to insert type annotations into the program AST.
Both \deeptyper{} and \lambdanet{} require type weaving, but their CSV files are in different formats.
Our type weaving program can be extended to support custom CSV formats.

\subsubsection{DeepTyper}\label{sec:approach-weaving-dt}

Each row of a \deeptyper{} CSV file represents a lexical token from the source program.
Rows with non-identifier tokens, such as keywords and symbols, contain two columns: the token text and the token type.
Rows with identifiers contain columns for the token text, token type, as well as the top five most likely types and their probabilities.

The \deeptyper{} implementation has a few limitations that we handle during type weaving.
First, the implementation uses regular expressions instead of a parser to tokenize JavaScript code.
This results in some tokens that are missing or incorrectly classified as identifiers.
Second, \deeptyper{} provides type predictions for every occurrence of an
identifier, so we must use only the predictions for declarations.
Finally, \deeptyper{} often predicts \texttt{complex} as a type; we do not believe it refers to a complex number type, so we replace it with \texttt{any}.

Our type weaving algorithm works as follows: as it traverses the program AST, if
it encounters a declaration node, it queries the CSV file for a type prediction.
However, the \deeptyper{} format does not record source location information,
and the token classification is brittle, so it is not straightforward to identify which rows are actually declarations and which rows should be skipped.
Our algorithm searches the CSV file for a short sequence of rows that corresponds to the declaration node in the AST.
This algorithm works well in practice, but does not handle optional parameters or statements that declare multiple variables.

\subsubsection{LambdaNet}\label{sec:approach-weaving-ln}

For each declaration, \lambdanet{} prints the source location of the identifier (start line, start column, end line, and end column), followed by the top five most likely types and their probabilities.
We modified \lambdanet{} to output in CSV format.

We observed that \lambdanet{} frequently predicts the following types: \texttt{Number}, \texttt{Boolean}, \texttt{String}, \texttt{Object}, and \texttt{Void}.
The first four are valid TypeScript types, but are non-primitive boxed types distinct from \texttt{number}, \texttt{boolean}, \texttt{string}, and \texttt{object}.
The TypeScript documentation strongly recommends using the lowercase type names,\footnote{\url{https://www.typescriptlang.org/docs/handbook/declaration-files/do-s-and-don-ts.html\#number-string-boolean-symbol-and-object}} so we normalize those types during type weaving.
Furthermore, \texttt{Void} is not a valid type, so we instead use \texttt{void}.
Finally, \lambdanet{} does not support generic types, but will predict them without type arguments, which is not valid in TypeScript.
While we cannot fix every generic type, we normalize \texttt{Array} to \texttt{any[]}, which is shorthand for \texttt{Array<any>}.

\enlargethispage{\baselineskip}
As our type weaving program traverses the program AST, if it encounters a
declaration node, it computes the node's source location information, and uses
that to query the CSV file for a type prediction.
However, the type annotation cannot be applied directly to the declaration node, as this modifies the AST and invalidates source location information.
Therefore, type weaving for \lambdanet{} occurs in two phases.
In the first phase, the traversal does not modify the AST, but saves the declaration node and type prediction in a map.
Then, in the second phase, type weaving iterates over the map and updates the AST.

\subsection{Type Checking}\label{sec:approach-type-checking}

In the final step, we run the TypeScript compiler to type check the migrated projects.
We run the compiler on each project, providing all the TypeScript input files as
arguments, and setting the following compiler flags:
\begin{description}
\item [\texttt{-{}-noEmit}] Type check only, do not emit JavaScript

\item [\texttt{-{}-esModuleInterop}] Improve handling of CommonJS and \es{} modules

\item [\texttt{-{}-moduleResolution node}] Explicitly set the module resolution strategy to Node.js

\item [\texttt{-{}-target es6}] Enable \es{} 6 features, which are used by some packages

\item [\texttt{-{}-lib es2021,dom}]  Include \es{} 2021 library definitions and browser DOM definitions
\end{description}
We do not set the \texttt{-{}-strict} flag, allowing the type checker
to be more lenient in certain situations, which we expect to already be a
significant challenge for automated type migration.
Furthermore, we ensure that package dependencies are properly included in our dataset so that the compiler can resolve them.

\section{Evaluation}\label{sec:eval}

\subsection{Dataset}\label{sec:eval-dataset}

\begin{table}[t]
\centering
\caption{Summary of dataset categories: number of packages, files, and lines of code.}\label{tbl:dataset}
\begin{tabularx}{0.675\textwidth}{ l r r r }
Dataset category & Packages & Files & Lines of code \\
\hline
DefinitelyTyped, no deps & 286 & 2,692 & 123,157 \\
DefinitelyTyped, with deps & 85 & 671 & 63,057 \\
Never typed, no deps & 102 & 255 & 20,729 \\
Never typed, with deps & 40 & 544 & 19,189 \\
\textbf{Overall} & 513 & 4,162 & 226,132 \\
\\
\end{tabularx}
\end{table}

\begin{figure}[t]
\centering
\includegraphics[width=0.8\textwidth]{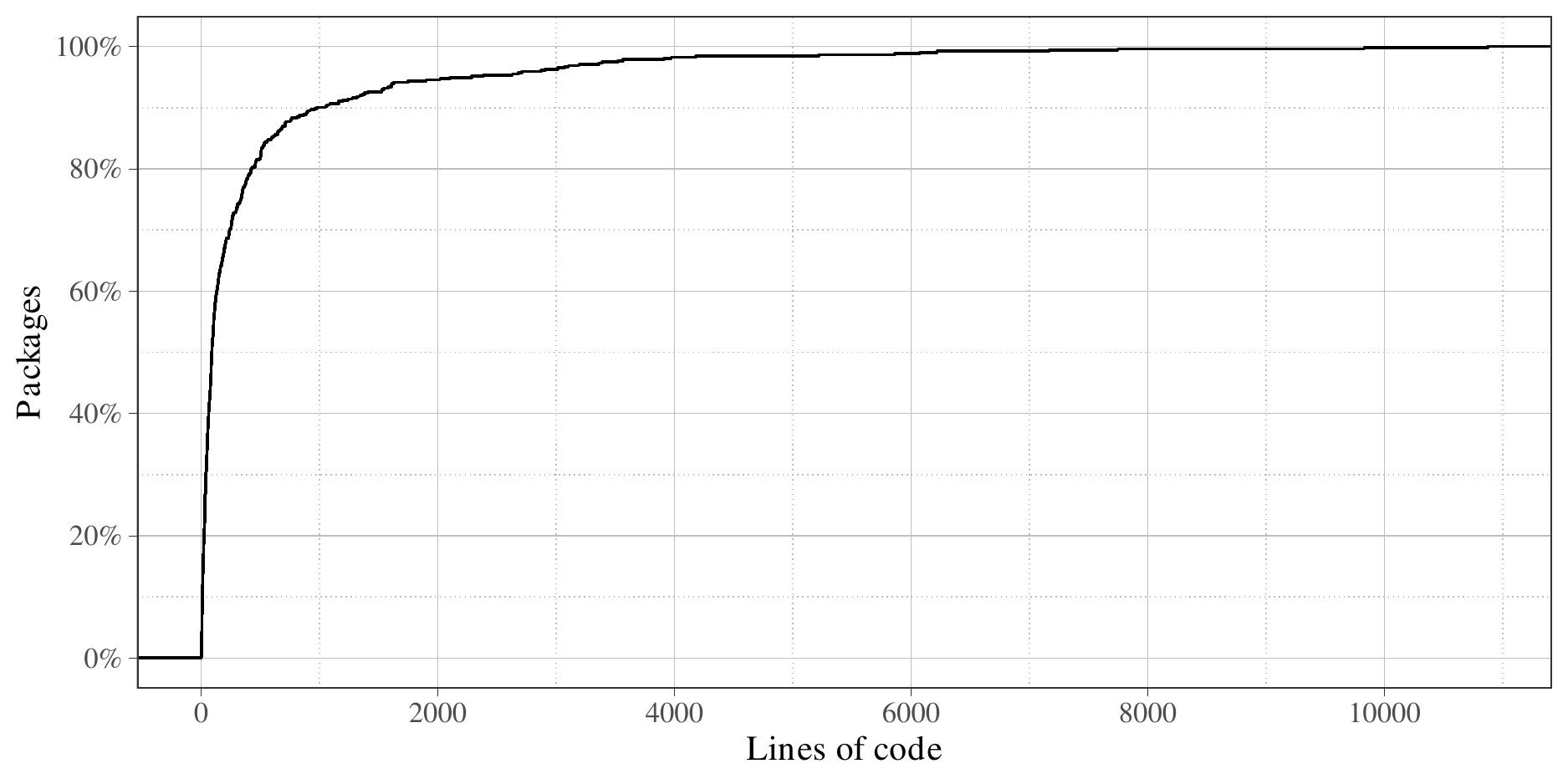}
\caption{Empirical cumulative distribution function of lines of code per package, over all datasets. The $x$-axis shows lines of code and the $y$-axis shows the proportion of packages with fewer than $x$ lines of code.}\label{fig:package-loc}
\end{figure}

Our dataset for evaluation consists of \totalpkgs{} JavaScript packages. To build this
dataset, we start from the top 1,000 most downloaded packages from the npm
Registry (as of August 2021) and narrow and clean as follows:
\begin{enumerate}

\item We add any transitive dependencies that are not in the original set of
packages, to ensure that the dataset is closed.

\item We try to fetch the original source code of every package, and
eliminate any package where this is not possible (\eg{}, the package did
not provide a repository URL or was deleted from GitHub). Fetching the
source helps ensure we are working with original code, and not compiled or
``minified'' JavaScript.

\item We remove packages that were built from a ``monorepo,'' \ie{}, a single
repository containing multiple packages that are published separately.
For example, the Babel JavaScript compiler has over 100
separate packages, but all share the same monorepo; fetching each source
package meant downloading the entire monorepo multiple times and including
unnecessary packages.

\item We remove packages that were not implemented in JavaScript, do not
contain code, or have more than 10,000 lines of code. The size limit helps
us avoid timeouts, and mostly excludes large toolchains and standard
libraries, such as the TypeScript compiler and \texttt{core-js} standard
library.

\item We remove testing code from every package. Tests frequently require
extra dependencies, and different frameworks set up the test environment in
different ways, which makes large-scale evaluation harder. To remove
testing code, we deleted directories named \texttt{test}, \texttt{tests},
\texttt{\_\_tests\_\_}, or \texttt{spec}, and files named \texttt{test.js},
\texttt{tests.js}, \texttt{test-*.js}, \texttt{*-test.js},
\texttt{*.test.js}, or \texttt{*.spec.js}.

\item Finally, we ensure that every package has \emph{no dependencies}, or
that \emph{all its dependencies are typed}, meaning the dependencies have
TypeScript type declaration (\texttt{.d.ts}) files available. (We do not
require that \emph{packages} are typed, but only that their
\emph{dependencies} are.)
This requirement is necessary because a JavaScript package can only be
imported into a TypeScript project if its interface has TypeScript type
declarations.
The DefinitelyTyped
repository\footnote{\url{https://github.com/DefinitelyTyped/DefinitelyTyped/}}
contains interface type declarations for many popular JavaScript packages,
and a handful of packages include their own.
We download type declarations of project dependencies and include them in
our dataset for evaluation purposes -- they are not used for type prediction.

\end{enumerate}

After filtering and cleaning our dataset, we classify every package with two
criteria: (1)~whether the package has type declarations available; and (2)~whether the package has dependencies.

\enlargethispage{1.5\baselineskip}
If a package has type declarations available, we say it is ``DefinitelyTyped'' and we can use its type
annotations as ground truth in our evaluation.
(However, there is evidence that some of these type annotations are
incorrect~\cite{feldthaus:ts-interface,kristensen:tsd,kristensen:tts,williams:ts-tpd}.)
Otherwise, we use the term ``never typed'': these packages \emph{have never
been type annotated} and thus no ground truth exists, so machine learning models
have never been evaluated on these packages before.
If a package has dependencies, we classify it as ``with deps'' (and from our
filtering, we know that every
dependency is typed); otherwise, we classify the package as ``no deps.''
Thus, there are four dataset categories; we list them in \cref{tbl:dataset}
along with the number of packages, files, and lines of code for each category.

\cref{fig:package-loc} is an empirical cumulative distribution function of the lines of code per package: the $x$-axis shows lines of code in a package and the $y$-axis shows the proportion of packages with fewer than $x$ lines of codes.
From the graph, we observe that approximately 90\% of packages have fewer than
1,000 lines of code, and approximately 95\% of packages have fewer than 2,000 lines of code.

\subsection{Success Rate of Type Checking}\label{sec:eval-type-checking}

\begin{table}[t]
\centering
\caption{Number and percentage of packages that type check.}\label{tbl:typecheck-packages}
\begin{tabularx}{0.935\textwidth}{ l r r r r r r r r r }
& \multicolumn{3}{c}{\deeptyper{}} & \multicolumn{3}{c}{\lambdanet{}} & \multicolumn{3}{c}{\incoder{}} \\
Dataset category & $\checkmark$ & Total & \% & $\checkmark$ & Total & \% & $\checkmark$ & Total & \% \\
\hline
DefinitelyTyped, no deps & 54 & 257 & 21.0 & 24 & 247 & 9.7 & 55 & 277 & 19.9 \\
DefinitelyTyped, with deps & 5 & 69 & 7.2 & 1 & 70 & 1.4 & 10 & 77 & 13.0 \\
Never typed, no deps & 31 & 95 & 32.6 & 11 & 87 & 12.6 & 25 & 100 & 25.0 \\
Never typed, with deps & 5 & 39 & 12.8 & 3 & 35 & 8.6 & 4 & 39 & 10.3 \\
\textbf{Overall} & 95 & 460 & 20.7 & 39 & 439 & 8.9 & 94 & 493 & 19.1 \\
\\
\end{tabularx}
\end{table}

\begin{figure}[t]
\centering
\includegraphics[width=0.8\textwidth]{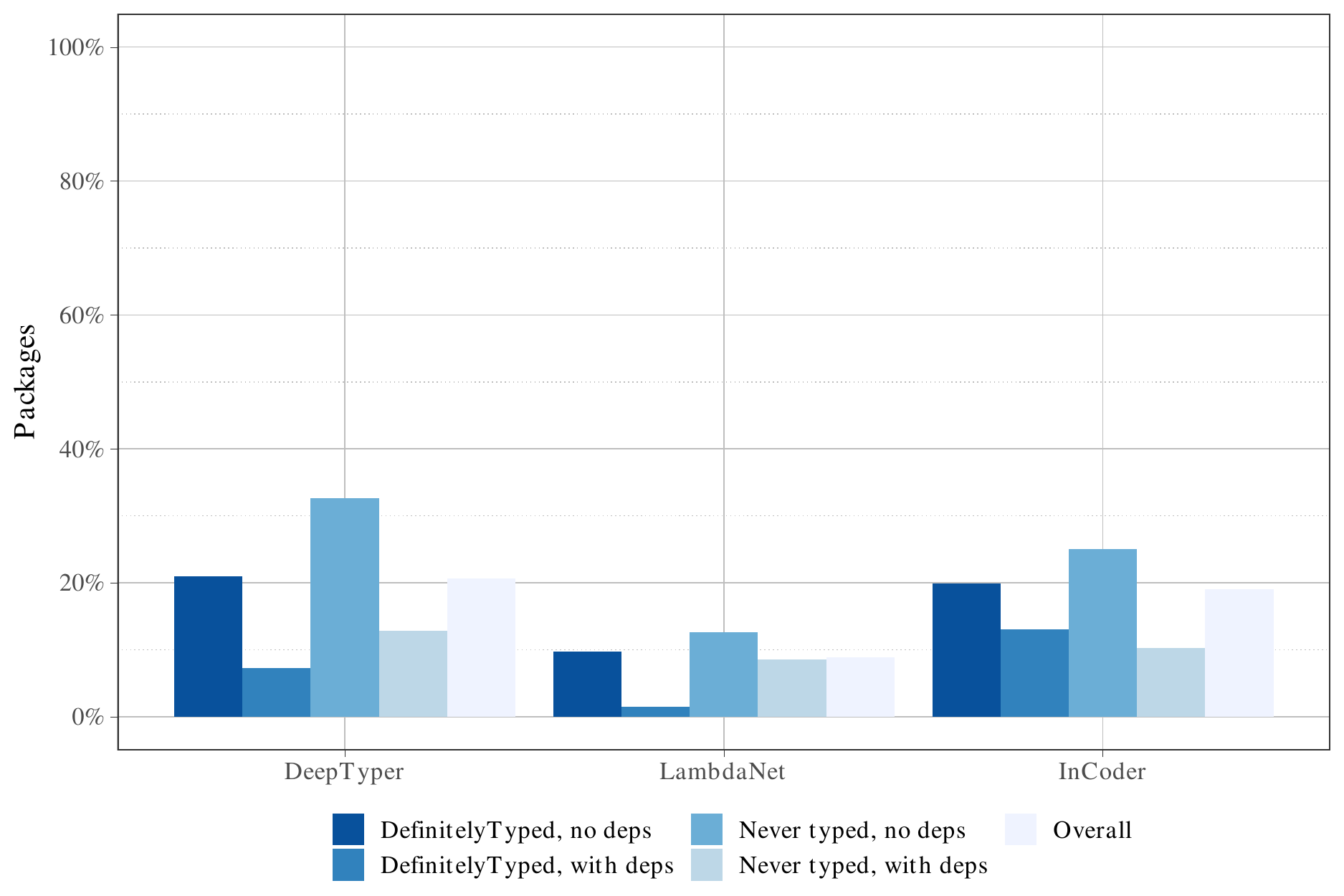}
\caption{Percentage of packages that type check.}\label{fig:typecheck-packages}
\end{figure}

\paragraph*{Do Migrated Packages Type Check?}

We first ask if entire packages type check after automated migration from
JavaScript to TypeScript. However, not all packages successfully translate to
TypeScript with every type migration tool; some packages cause the type
migration tool to time out or error. Thus, we report the success rate of
type checking as a fraction of the packages that successfully translate to
TypeScript.

\Cref{tbl:typecheck-packages} and \cref{fig:typecheck-packages} show the
fraction of packages that type check with each tool. We observe that
\deeptyper{} and \incoder{} perform similarly (20\% success rate), and
\lambdanet{} performs worse (9\% success rate).
Across all tools, packages without dependencies type check at a higher rate
than packages with dependencies.

These package-level type checking results are disappointing -- but this is
a very high standard to meet. Even a single incorrect type annotation causes
the entire package to fail. Therefore, we next consider a finer-grained metric
that is still useful.

\paragraph*{How Many Files are Error Free?}

\begin{table}[p]
\centering
\caption{Number and percentage of files with no compilation errors.}\label{tbl:files-no-errors}
\begin{tabularx}{0.985\textwidth}{ l r r r r r r r r r }
& \multicolumn{3}{c}{\deeptyper{}} & \multicolumn{3}{c}{\lambdanet{}} & \multicolumn{3}{c}{\incoder{}} \\
Dataset category & $\checkmark$ & Total & \% & $\checkmark$ & Total & \% & $\checkmark$ & Total & \% \\
\hline
DefinitelyTyped, no deps & 414 & 1,010 & 41.0 & 474 & 1,638 & 28.9 & 1,689 & 2,401 & 70.3 \\
DefinitelyTyped, with deps & 195 & 384 & 50.8 & 169 & 504 & 33.5 & 312 & 547 & 57.0 \\
Never typed, no deps & 95 & 205 & 46.3 & 63 & 229 & 27.5 & 101 & 235 & 43.0 \\
Never typed, with deps & 42 & 121 & 34.7 & 25 & 534 & 4.7 & 467 & 527 & 88.6 \\
\textbf{Overall} & 746 & 1,720 & 43.4 & 731 & 2,905 & 25.2 & 2,569 & 3,710 & 69.2 \\
\\
\end{tabularx}
\end{table}

\begin{figure}[p]
\centering
\includegraphics[width=0.8\textwidth]{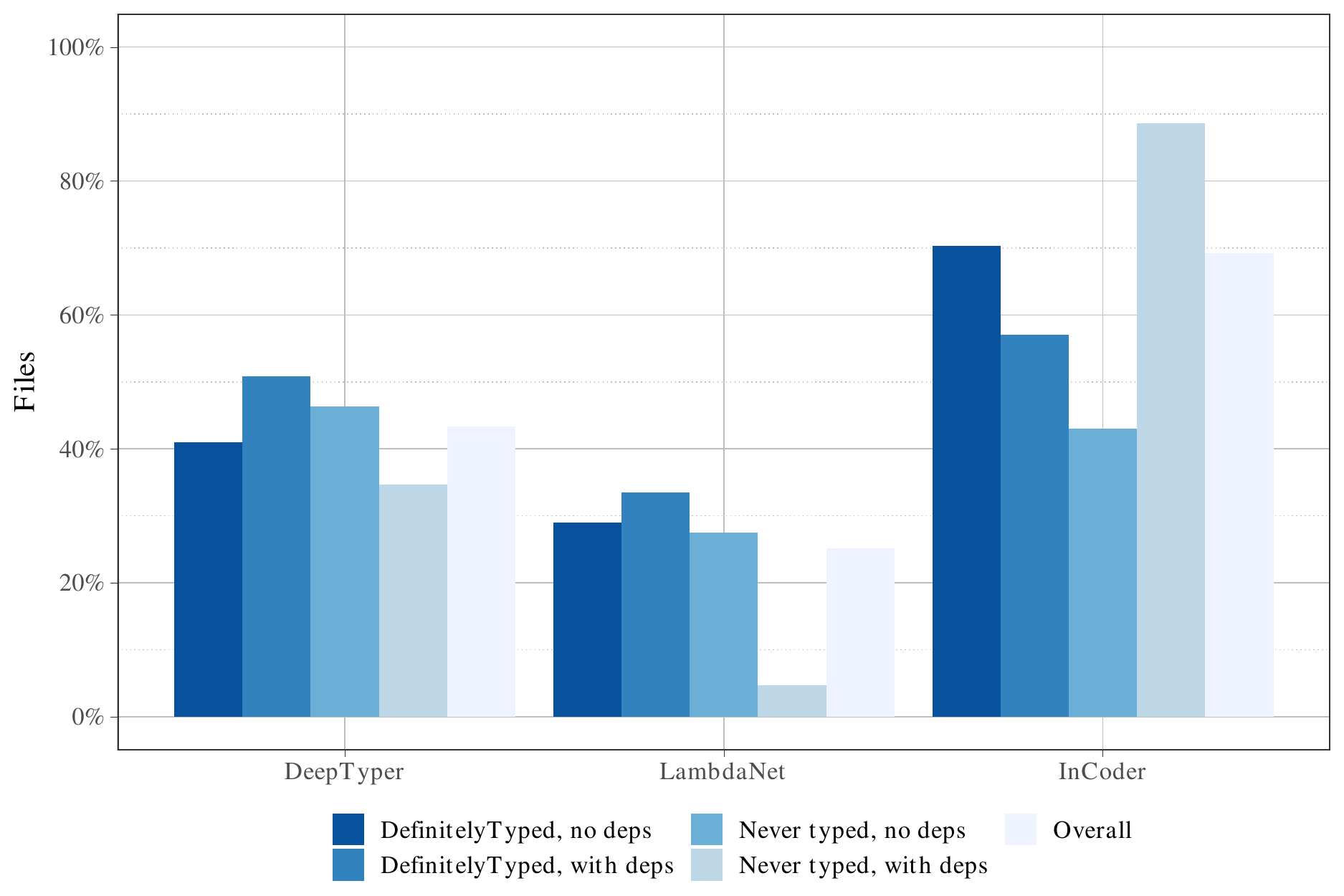}
\caption{Percentage of files with no compilation errors.}\label{fig:files-no-errors}
\end{figure}

\begin{figure}[p]
\centering
\includegraphics[width=0.8\textwidth]{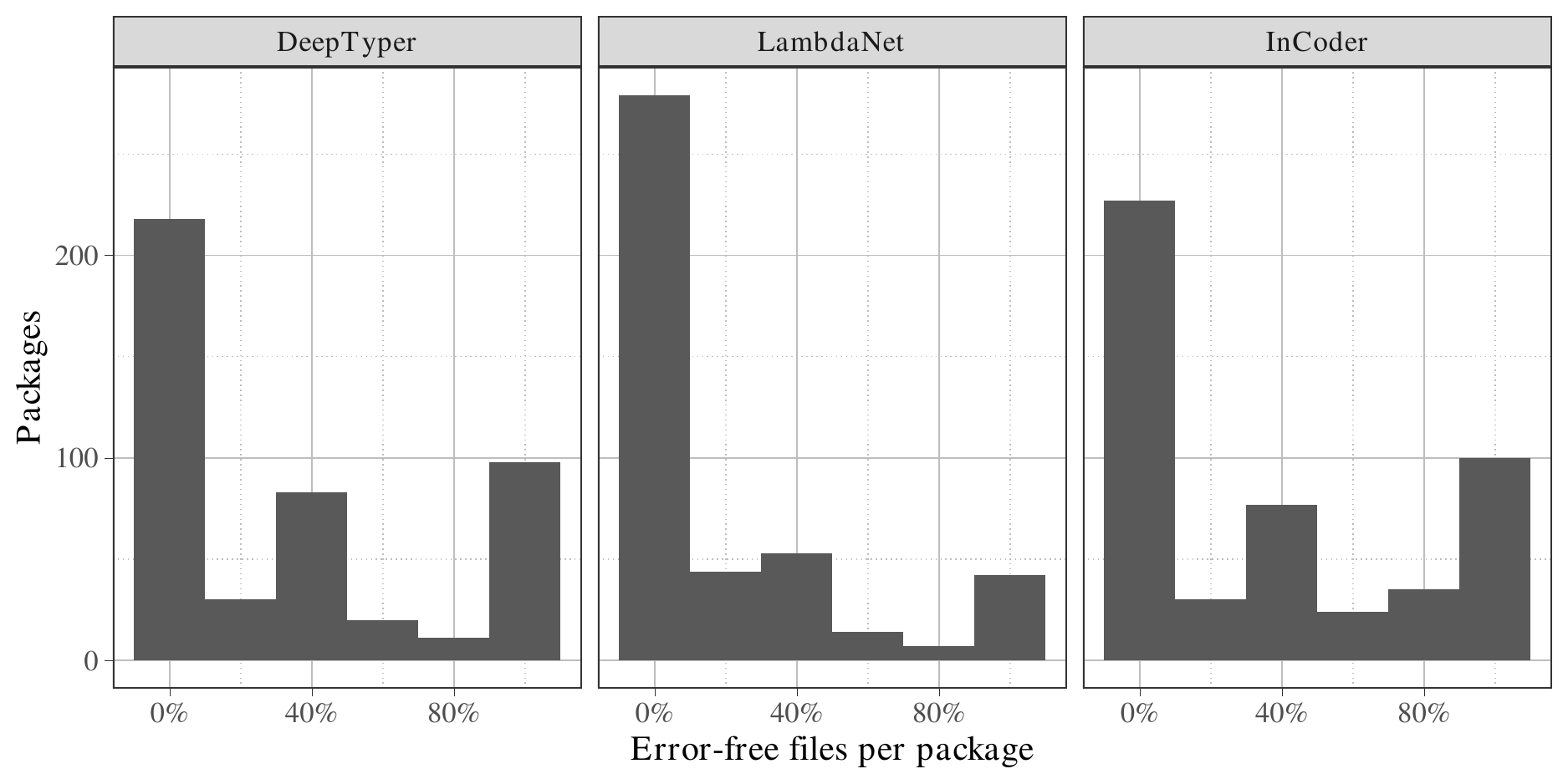}
\caption{Number of packages vs.\ percentage of error-free files per package.}\label{fig:errorfree-per-package}
\end{figure}

As an alternate measure, we look at the percentage of \emph{files with no
compilation errors}. Instead of a binary pass/fail outcome, this gives
us a more fine-grained result for a package. We motivate this metric by
observing that TypeScript files are modules with explicit imports and exports.
If a file type checks without errors, then it is using all of its internal and
imported types consistently.
Thus, when triaging type errors, a programmer may (temporarily) set these
files aside and focus on the files with compilation errors.
However, the programmer may later need to return to a file with no type errors and
adjust its type annotations, for example, if a consumer of that file expects a
different interface.
We give examples of this in our case studies, specifically \cref{sec:cs-bad-annotation} and \cref{sec:case-study-type-guard}.

\cref{tbl:files-no-errors} and \cref{fig:files-no-errors} present the fraction
of files with no compilation errors. The results are more encouraging: using
\incoder{}, 69\% of files are error free. With these results, it is
not clear that packages without dependencies outperform packages with
dependencies.
Finally, in \cref{fig:errorfree-per-package} we calculate the percentage
of error-free files for each package, and plot histograms of the
distribution. Across all tools, most packages have type errors in most or all
files.

\paragraph*{What Percentage of Type Annotations Are Trivial?}

\begin{figure}[t]
\centering
\includegraphics[width=0.8\textwidth]{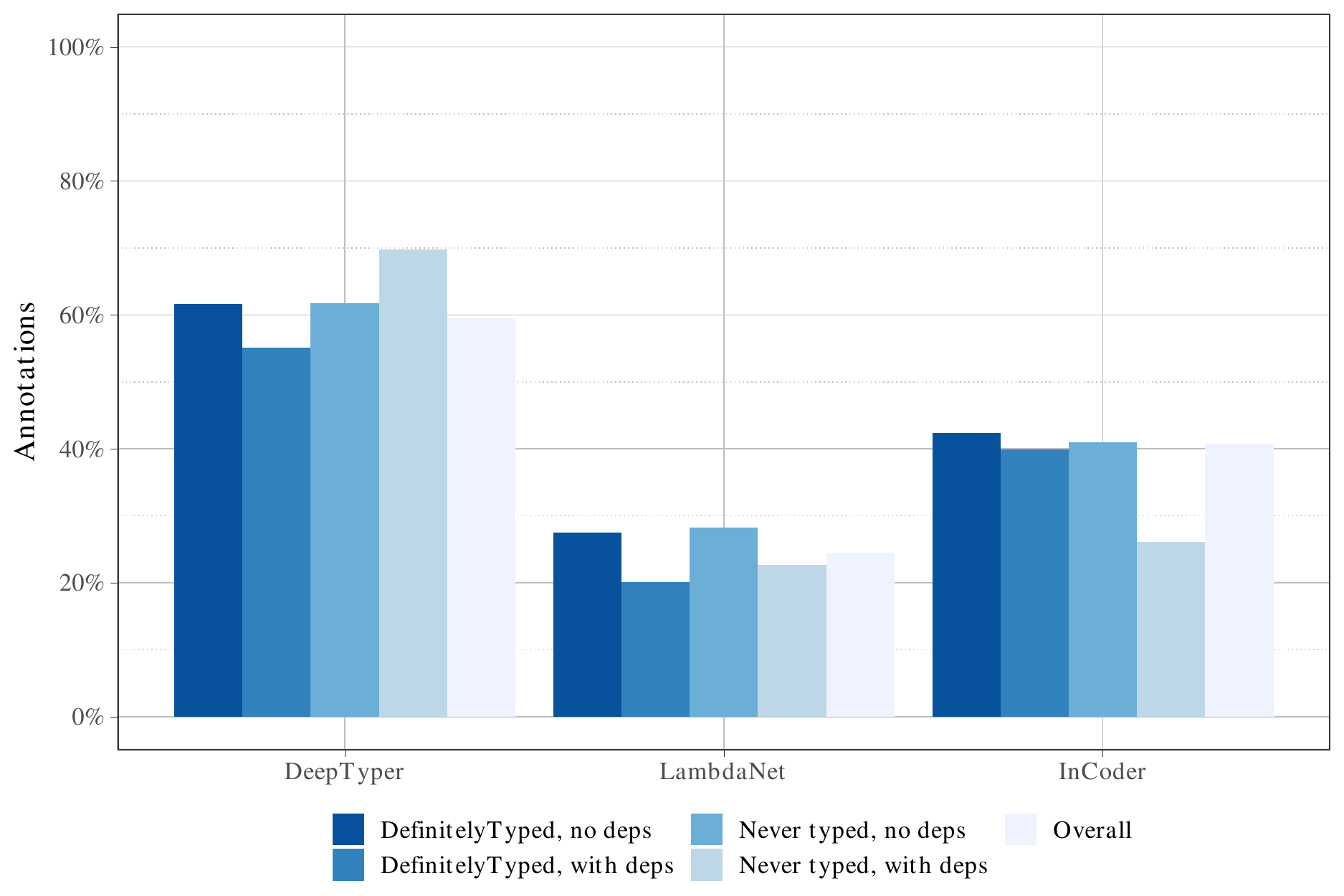}
\caption{Percentage of annotations that are \texttt{any},
\texttt{any[]}, or \texttt{Function}, in files with no
errors.}\label{fig:errorfree-files-anys}
\end{figure}

Next, we examine what percentage of type annotations, \emph{within the error-free
files}, are trivial, \ie, what  percentage are
\texttt{any}, \texttt{any[]} (array of \texttt{any}s), or
\texttt{Function} (function that accepts any arguments
and returns anything).
These annotations can hide type errors and allow more
code to type check; however, they provide little value to the programmer.

\cref{fig:errorfree-files-anys} shows the percentage of trivial type
annotations within error-free files.
\deeptyper{} produces the most (about 60\%),
\lambdanet{} produces the least (about 25\%), while \incoder{} is in between
(about 40\%).

Comparing to the percentage of files with no compilation errors
(\cref{fig:files-no-errors}), \deeptyper{} produces more type-correct code than
\lambdanet{}, but it also generates more trivial type annotations. \incoder{}
produces the most type-correct code, while generating a moderate
percentage of trivial type annotations.

\paragraph*{Do Migrated Types Match Human-Written Types (When Available)?}

\begin{table}[t]
\centering
\caption{Accuracy of type annotations, compared to non-\texttt{any} ground truth.}\label{tbl:accuracy}
\begin{tabularx}{0.955\textwidth}{ l r r r r r r r r r }
& \multicolumn{3}{c}{\deeptyper{}} & \multicolumn{3}{c}{\lambdanet{}} & \multicolumn{3}{c}{\incoder{}} \\
Dataset category & $\checkmark$ & Total & \% & $\checkmark$ & Total & \% & $\checkmark$ & Total & \% \\
\hline
DefinitelyTyped, no deps & 90 & 241 & 37.3 & 116 & 259 & 44.8 & 40 & 123 & 32.5 \\
DefinitelyTyped, with deps & 27 & 123 & 22.0 & 41 & 119 & 34.5 & 11 & 64 & 17.2 \\
\textbf{Overall} & 117 & 364 & 32.1 & 157 & 378 & 41.5 & 51 & 187 & 27.3 \\
\\
\end{tabularx}
\end{table}

\begin{figure}[t]
\centering
\includegraphics[width=0.8\textwidth]{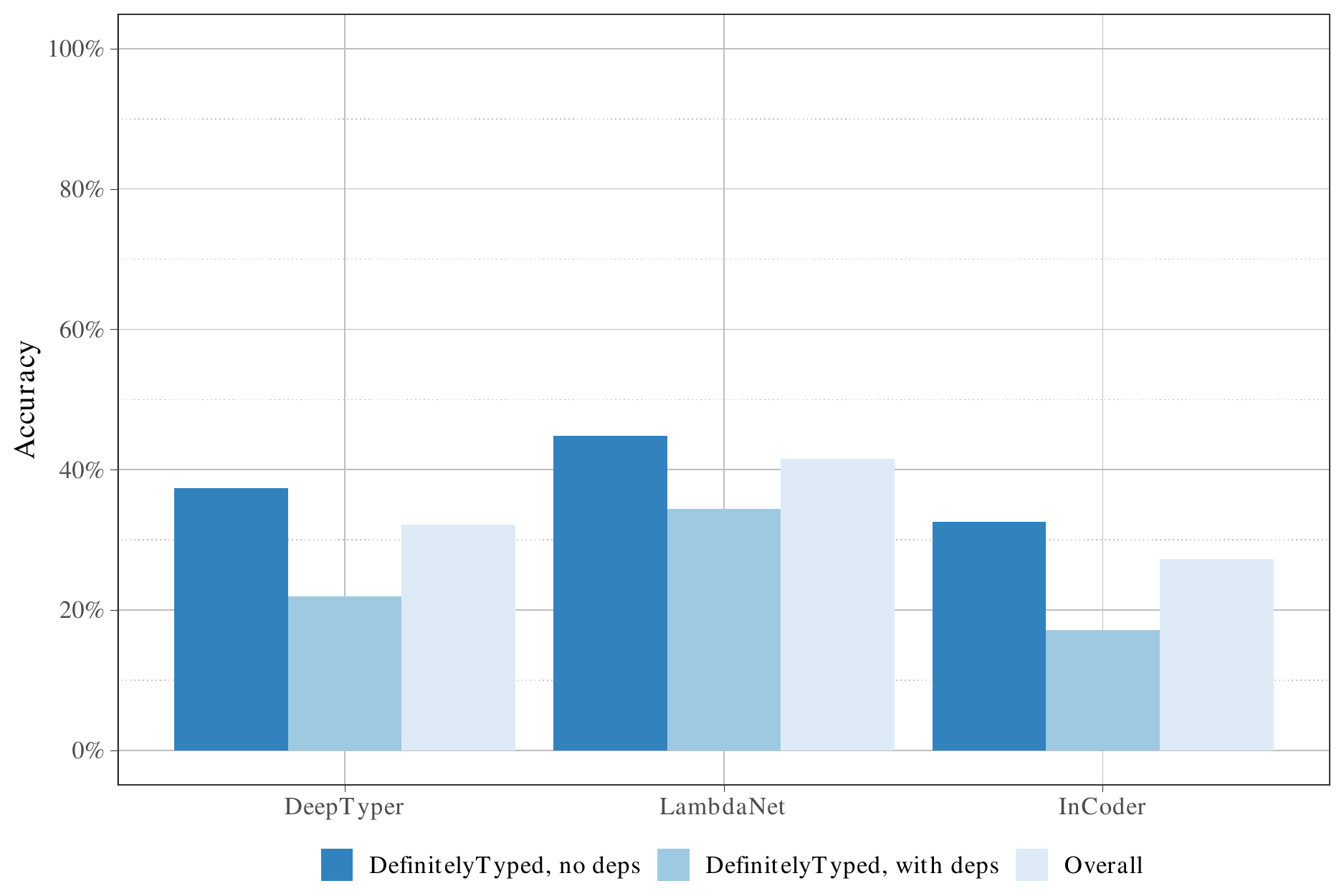}
\caption{Accuracy of type annotations, compared to non-\texttt{any} ground truth.}\label{fig:accuracy}
\end{figure}

Since our dataset is constructed from JavaScript packages instead of TypeScript
packages, we do not have fully type-annotated files as our ground truth;
instead, we use declaration files provided by the DefinitelyTyped repository
or package author. We configure the TypeScript compiler to
emit declarations during type checking, which it can do even
if the whole package does not type check.
Thus, we can compare handwritten, ground truth declarations against declarations generated from migrated packages.

We extract function signatures from declaration files and only compare a signature if it is in both the ground truth and generated declaration.
We compare the function parameter types and return types one-to-one, ignoring modifiers (\eg{}, \texttt{readonly}), and we require an exact string match (\ie{} \texttt{string | number} and \texttt{number | string} are considered different types).
Following the literature, we skip a comparison if the ground truth is the \texttt{any} annotation.

\enlargethispage{1.9\baselineskip}
Our results are presented in \cref{tbl:accuracy} and \cref{fig:accuracy}.
We observe that accuracy is better for packages without dependencies.
Additionally, our results follow the same pattern in prior work, where
\lambdanet{} has better accuracy than \deeptyper{}, despite performing worse in
our other metrics.

\paragraph*{How Many Errors Occur in Each Package?}

\begin{figure}[t]
\centering
\includegraphics[width=0.8\textwidth]{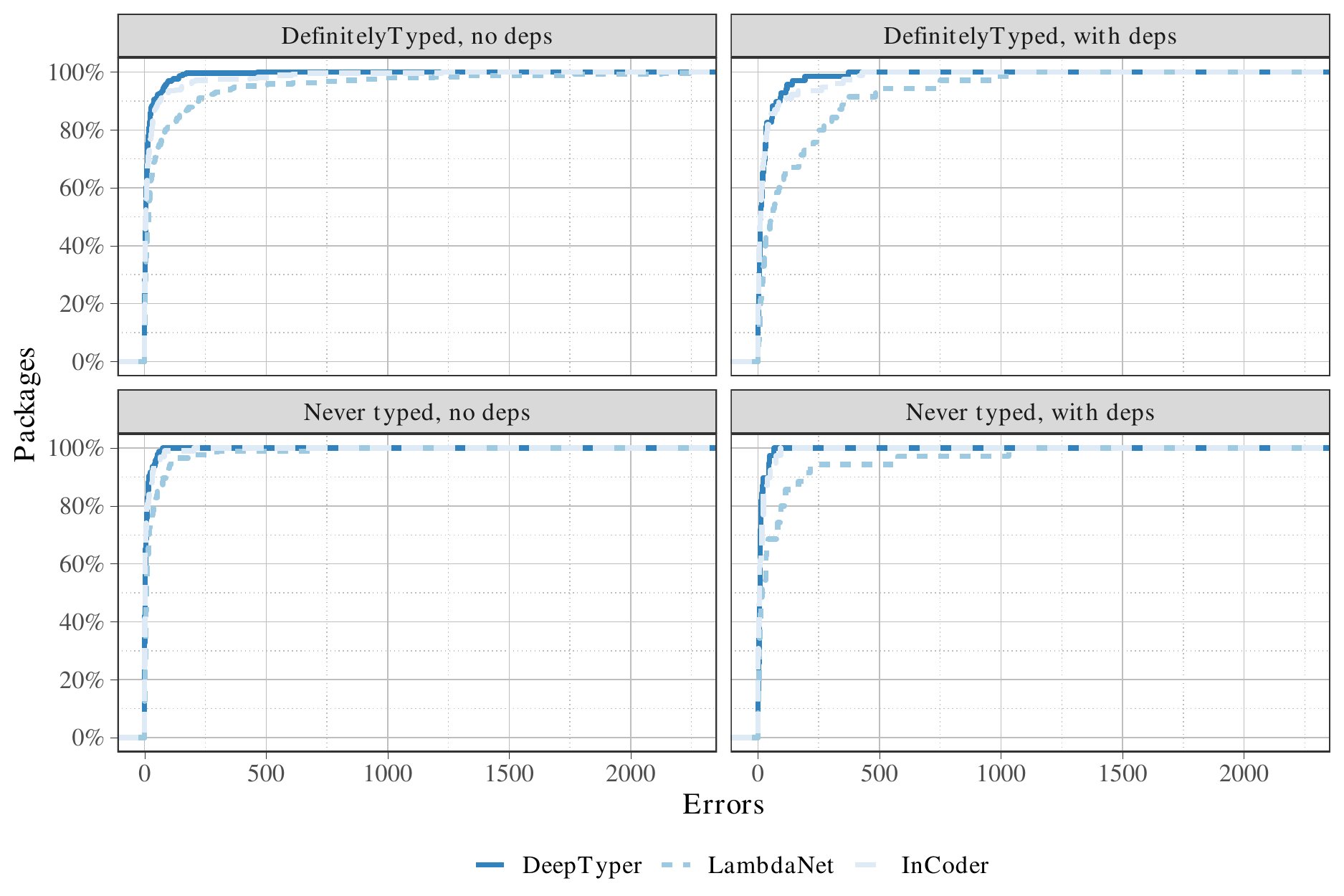}
\caption{Empirical cumulative distribution function of errors. The $x$-axis shows the number of errors and the $y$-axis shows the proportion of packages with fewer than $x$ errors.}\label{fig:error_cdf}
\end{figure}

\cref{fig:error_cdf} shows an empirical cumulative distribution function of
errors: the $x$-axis shows the number of errors and the $y$-axis shows the proportion of packages with fewer than $x$ errors.
For example, when migrating the ``DefinitelyTyped, with deps'' dataset with \lambdanet{}, approximately 80\% of packages have fewer than 250 errors each.
Additionally, all of \deeptyper{}'s packages and almost all of \incoder{}'s packages have fewer than 500 errors each.

\subsection{Error Analysis}\label{sec:eval-error-analysis}

\begin{figure}[t]
\centering
\includegraphics[width=0.8\textwidth]{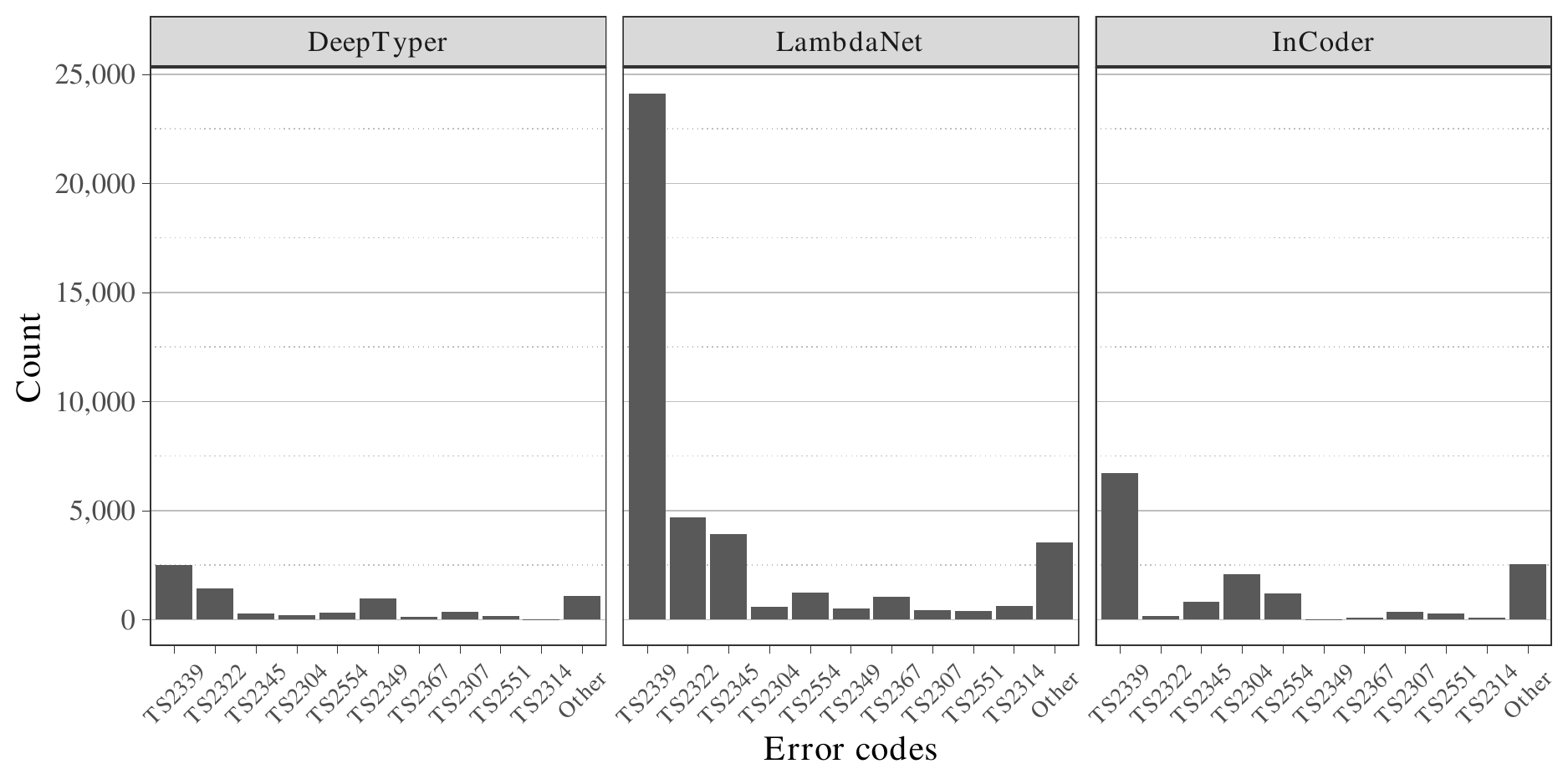}
\caption{Distribution of the top 10 most common error codes, over all datasets.}\label{fig:error-codes}
\end{figure}

\begin{table}[t]
\centering
\caption{The top 10 most common error codes.}\label{tbl:error-codes}
\begin{tabularx}{\textwidth}{ l X r r r }
Error code & Message & \deeptyper{} & \lambdanet{} & \incoder{} \\
\hline
TS2339 & Property '\{0\}' does not exist on type '\{1\}.' & 2,510 & 24,123 & 6,742 \\
TS2322 & Type '\{0\}' is not assignable to type '\{1\}'. & 1,429 & 4,709 & 176 \\
TS2345 & Argument of type '\{0\}' is not assignable to parameter of type '\{1\}'. & 304 & 3,930 & 828 \\
TS2304 & Cannot find name '\{0\}'. & 217 & 598 & 2,094 \\
TS2554 & Expected \{0\} arguments, but got \{1\}. & 330 & 1,234 & 1,200 \\
TS2349 & This expression is not callable. & 977 & 529 & 26 \\
TS2367 & This condition will always return '\{0\}' since the types '\{1\}' and '\{2\}' have no overlap. & 145 & 1,046 & 110 \\
TS2307 & Cannot find module '\{0\}' or its corresponding type declarations. & 375 & 432 & 371 \\
TS2551 & Property '\{0\}' does not exist on type '\{1\}'. Did you mean '\{2\}'? & 187 & 389 & 296 \\
TS2314 & Generic type '\{0\}' requires \{1\} type argument(s). & 1 & 630 & 95 \\
Other &  & 1,089 & 3,528 & 2,557 \\
\textbf{Total} &  & 7,564 & 41,148 & 14,495 \\
\\
\end{tabularx}
\end{table}

We now consider the kinds of errors that arise during migration. Every
TypeScript compiler error has an associated code,\footnote{\url{https://github.com/Microsoft/TypeScript/blob/v4.9.3/src/compiler/diagnosticMessages.json}} making categorization straightforward.
\cref{fig:error-codes} summarizes the top 10 most common errors and \cref{tbl:error-codes} provides the corresponding messages.

\enlargethispage{\baselineskip}
Most of the errors relate to types. These errors are the following:
a property not existing on a type~(TS2339 and TS2551); an assignment with
mismatched types~(TS2322); a function call with mismatched parameter and
argument types~(TS2345); calling a function that was assigned a non-function
type annotation~(TS2349); and a conditional that compares values from different
types~(TS2367).
TS2314 refers to a generic type that was not provided type arguments; this is caused by \deeptyper{} and \lambdanet{} not fully supporting generic types.

The remaining errors are not directly related to types.
TS2304 refers to an unknown name, which may not necessarily be a type.
TS2554 is emitted because TypeScript requires the number of call arguments
to match the number of function parameters, but JavaScript does not.
TS2339 includes cases where an empty object is initialized by setting its properties,
but TypeScript requires that the object's properties are declared in its type.
Finally, TS2307 indicates that the \es{} module conversion produced incorrect code.

\subsection{\es{} Module Conversion}\label{sec:eval-cjs-vs-es}

\begin{table}[p]
\centering
\caption{Percentage of packages that type check, before and after \es{} module conversion.}\label{tbl:typecheck-comp}
\begin{tabularx}{0.875\textwidth}{ l r r r r r r }
& \multicolumn{2}{c}{\deeptyper{}} & \multicolumn{2}{c}{\lambdanet{}} & \multicolumn{2}{c}{\incoder{}} \\
Dataset category & Before & After & Before & After & Before & After \\
\hline
DefinitelyTyped, no deps & 25.3 & 21.0 & 8.8 & 9.7 & 21.3 & 19.9 \\
DefinitelyTyped, with deps & 11.6 & 7.2 & 2.9 & 1.4 & 14.3 & 13.0 \\
Never typed, no deps & 34.7 & 32.6 & 11.5 & 12.6 & 26.0 & 25.0 \\
Never typed, with deps & 28.2 & 12.8 & 8.8 & 8.6 & 23.1 & 10.3 \\
\textbf{Overall} & 25.4 & 20.7 & 8.4 & 8.9 & 21.3 & 19.1 \\
\\
\end{tabularx}
\end{table}

\begin{table}[p]
\centering
\caption{Percentage of files with no compilation errors, before and after \es{} module conversion.}\label{tbl:errorfree-comp}
\begin{tabularx}{0.875\textwidth}{ l r r r r r r }
& \multicolumn{2}{c}{\deeptyper{}} & \multicolumn{2}{c}{\lambdanet{}} & \multicolumn{2}{c}{\incoder{}} \\
Dataset category & Before & After & Before & After & Before & After \\
\hline
DefinitelyTyped, no deps & 45.3 & 41.0 & 27.9 & 28.9 & 68.3 & 70.3 \\
DefinitelyTyped, with deps & 50.0 & 50.8 & 24.3 & 33.5 & 51.6 & 57.0 \\
Never typed, no deps & 48.8 & 46.3 & 24.9 & 27.5 & 42.6 & 43.0 \\
Never typed, with deps & 64.5 & 34.7 & 8.6 & 4.7 & 11.2 & 88.6 \\
\textbf{Overall} & 48.1 & 43.4 & 23.5 & 25.2 & 56.1 & 69.2 \\
\\
\end{tabularx}
\end{table}

\begin{table}[p]
\centering
\caption{Accuracy of type annotations, before and after \es{} module conversion.}\label{tbl:accuracy-comp}
\begin{tabularx}{0.875\textwidth}{ l r r r r r r }
& \multicolumn{2}{c}{\deeptyper{}} & \multicolumn{2}{c}{\lambdanet{}} & \multicolumn{2}{c}{\incoder{}} \\
Dataset category & Before & After & Before & After & Before & After \\
\hline
DefinitelyTyped, no deps & 35.2 & 37.3 & 44.4 & 44.8 & 34.0 & 32.5 \\
DefinitelyTyped, with deps & 19.8 & 22.0 & 27.0 & 34.5 & 19.2 & 17.2 \\
\textbf{Overall} & 28.7 & 32.1 & 38.9 & 41.5 & 29.1 & 27.3 \\
\\
\end{tabularx}
\end{table}

\begin{figure}[p]
\centering
\includegraphics[width=0.8\textwidth]{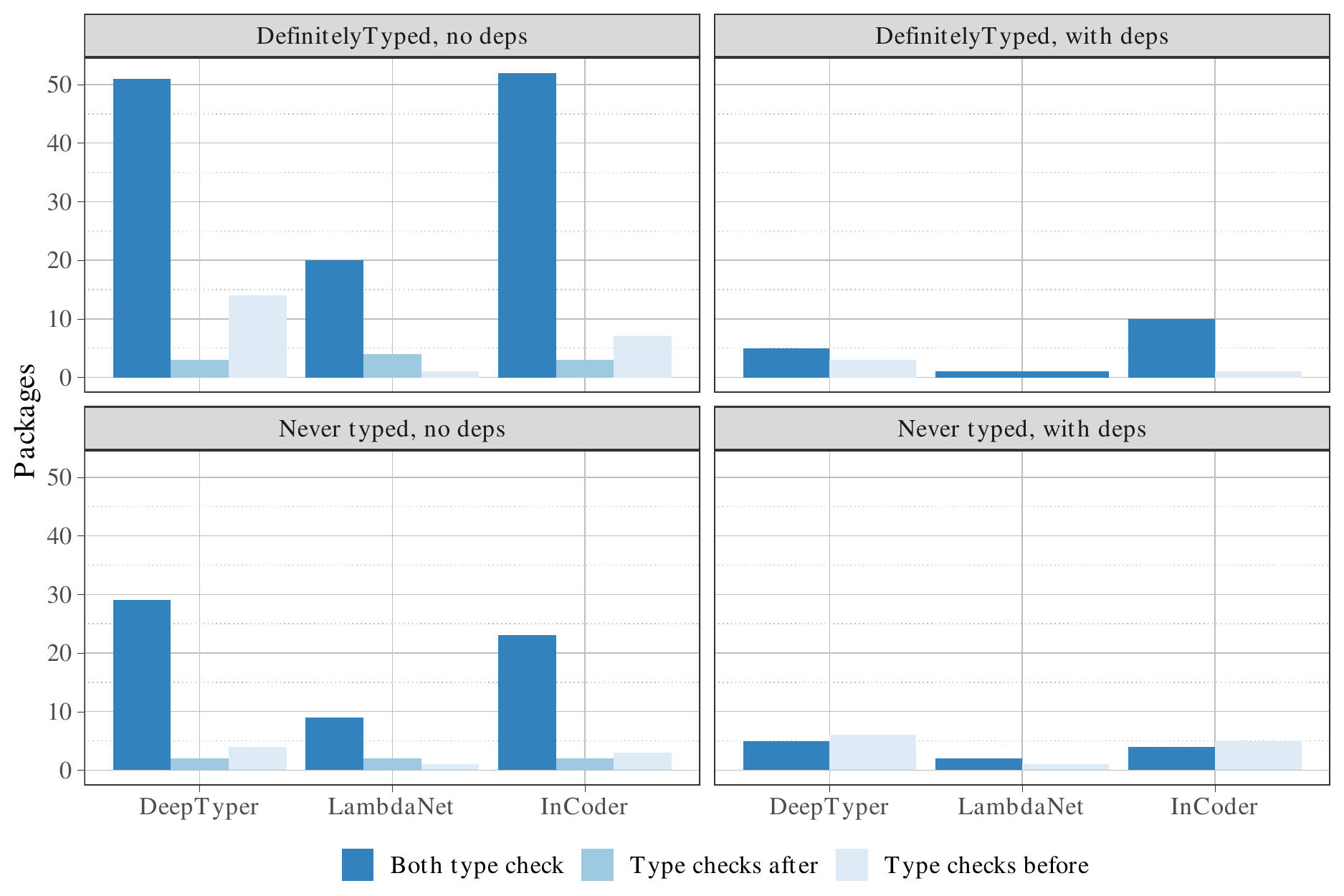}
\caption{Packages that type check before or after \es{} module conversion.}\label{fig:typecheck-comp}
\end{figure}

Recall that we described an optional step before running the evaluation: converting packages to use \es{} modules.
In this section, we re-run our evaluation -- generating types, weaving types, and type checking -- to compare the results before and after the conversion step.
Specifically, we examine the percentage of packages that type check (\cref{tbl:typecheck-comp}), the percentage of files with no errors (\cref{tbl:errorfree-comp}), and accuracy (\cref{tbl:accuracy-comp}).
However, this is not a direct comparison between CommonJS and \es{} modules, as some of the original packages were already using \es{} modules. Furthermore, the conversion affected a small handful of packages: some packages successfully migrated to TypeScript after the conversion but failed before, and the inverse was true for other packages.

From \cref{tbl:typecheck-comp}, we observe that the \es{} module conversion makes fewer packages type check for \deeptyper{} and \incoder{}, but slightly improves the results for \lambdanet{}.
In \cref{fig:typecheck-comp}, we examine packages that type checked before or after the \es{} module conversion; packages that never type checked were excluded.
In general, if a package type checked before the conversion, it likely type checked after the conversion.
However, if a package failed to type check before the conversion, it was unlikely to type check afterwards; in fact, this never happened for a package with dependencies.

\cref{tbl:errorfree-comp} compares the percentage of files with no compilation errors.
The conversion improves the results for \lambdanet{} and \incoder{}, but makes the results worse for \deeptyper{}.
One dramatic result is the change for \incoder{} and the ``never typed, with deps'' dataset, where the \es{} module conversion results in 89\% of files type checking, when it was only 11\% before.
The difference is caused by a single package, \texttt{regenerate-unicode-properties}, which has over 400 files.
With CommonJS modules, each file produces an error; however, with \es{} modules,
those files type check successfully.

Finally, \cref{tbl:accuracy-comp} compares the accuracy of type annotations, before and after the \es{} module conversion.
Recall that for accuracy, we compare type annotations against the ground truth of handwritten TypeScript declaration files; these are the ``DefinitelyTyped'' datasets.
Accuracy improves for \deeptyper{} and \lambdanet{}, but worsens slightly for \incoder{}.

\subsection{Case Studies}\label{sec:case-study}

In this section, we examine how the models performed on four packages, whether
the packages type check, and what steps are left to migrate the packages to
TypeScript.

\subsubsection{Error Message Does Not Refer to Incorrect Type Annotation}\label{sec:cs-error-message}

\begin{figure}[t]
\centering
\begin{lstlisting}
// Original
const handlePreserveConsecutiveUppercase =#\label{line:decamelize-original}#
  (decamelized, separator) => {#\label{line:decamelize-function}#
    // code omitted and simplified...
    return decamelized.replace(#\label{line:decamelize-replace}#
      /([A-Z]+)([A-Z][a-z]+)/gu,#\label{line:decamelize-regex}#
      (_, #\$#1, #\$#2) => #\$#1 + separator + #\$#2.toLowerCase(),#\label{line:decamelize-concat}#
    );
}

// Elsewhere in the package; str and sep are both strings
return handlePreserveConsecutiveUppercase(str, sep);#\label{line:decamelize-call}#

// DeepTyper solution
const handlePreserveConsecutiveUppercase: string =#\label{line:decamelize-dt}#
  (decamelized: string, separator: string) => { ... }#\label{line:decamelize-dt-func}#

// LambdaNet solution
const handlePreserveConsecutiveUppercase: Function =#\label{line:decamelize-ln}#
  (decamelized: string, separator: number) => { ... }#\label{line:decamelize-ln-func}#
    \end{lstlisting}
\caption{The \texttt{handlePreserveConsecutiveUppercase} function adapted
from the \texttt{decamelize} package. The \deeptyper{} and \lambdanet{}
solutions are also shown.}\label{fig:cs-decamelize}
\end{figure}

\texttt{decamelize} is a package for converting strings in camel case to lowercase.\footnote{\url{https://www.npmjs.com/package/decamelize}}
It is in the ``DefinitelyTyped, no deps'' dataset, as it ships with a
\texttt{.d.ts} declaration file and has no dependencies.
\cref{fig:cs-decamelize} shows a simplified version of the function
\texttt{handle\-Preserve\-Consecutive\-Uppercase}.
This function is not exported, thus there are no programmer-written type
annotations.
\cref{line:decamelize-function} uses JavaScript's arrow function notation
to define a function that takes two arguments, and assigns it to the constant on \cref{line:decamelize-original}.
Elsewhere in the package (\cref{line:decamelize-call} in the listing), the helper function is called with \texttt{str} and \texttt{sep} string arguments.

\enlargethispage{\baselineskip}
A programmer inspecting the function can reason that \cref{line:decamelize-replace} is a call to a string method that uses a regular expression on \cref{line:decamelize-regex} to replace text in \texttt{decamelized} with the result on \cref{line:decamelize-concat}, where \texttt{separator} is concatenated with the regular expression match. Therefore, both the \texttt{decamelized} and \texttt{separator} parameters on \cref{line:decamelize-function} should be annotated as \texttt{string}.

The \deeptyper{} solution is listed on \cref{line:decamelize-dt}: it correctly
annotates both function parameters as \texttt{string}, but incorrectly annotates
\texttt{handle\-Preserve\-Consecutive\-Uppercase} as \texttt{string}. The
compiler emits errors for \cref{line:decamelize-dt-func,line:decamelize-call},
because \cref{line:decamelize-call} is attempting to call a non-function, and
\cref{line:decamelize-dt-func} is attempting to assign a function to a
non-function variable. However, the fix must be applied to the annotation on
line \cref{line:decamelize-dt}.

\subsubsection{Incorrect Type Annotation Can Type Check Successfully}\label{sec:cs-bad-annotation}

\begin{figure}[t]
\centering
\begin{lstlisting}
// Original
export const write =#\label{line:ieee754-original}#
  function (buffer, value, offset, isLE, mLen, nBytes) { ... }

// Ground truth signature
export function write(#\label{line:ieee754-groundtruth}#
  buffer: Uint8Array, value: number, offset: number, isLE: boolean,
  mLen: number, nBytes: number): void;

// DeepTyper solution
export const write: void = function (#\label{line:ieee754-dt}#
  buffer: Buffer, value: number, offset: number, isLE: number,#\label{line:ieee754-dt-isle}#
  mLen: number, nBytes: number) { ... }

// InCoder solution
export const write = function (#\label{line:ieee754-ic}#
  buffer: Buffer, value: any,#\label{line:ieee754-ic-value}# offset: number, isLE: boolean,
  mLen: number, nBytes: number) { ... }
    \end{lstlisting}
\caption{The \texttt{write} function adapted from the \texttt{ieee754}
package. The ground truth signature is also shown, along with the
\deeptyper{} and \incoder{} solutions.}\label{fig:cs-ieee754}
\end{figure}

The \lambdanet{} solution on \cref{line:decamelize-ln} correctly annotates
\texttt{handle\-Preserve\-Consecutive\-Uppercase} as \texttt{Function}, but it
incorrectly annotates the \texttt{separator} parameter on
\cref{line:decamelize-ln-func} as \texttt{number}.
We would expect the compiler to emit a type error, since
\cref{line:decamelize-call} calls the function with string arguments.
However, the code type checks successfully, because the generic
\texttt{Function} type on \cref{line:decamelize-ln} accepts any number of arguments of any type.
The \texttt{Function} type annotation is similar to \texttt{any}, in that it enables more code to type check, but at the cost of fewer type guarantees.

Another example of this problem is the \texttt{ieee754} package, which reads and writes floating point numbers to and from buffers.\footnote{\url{https://www.npmjs.com/package/ieee754}}
It is categorized as ``DefinitelyTyped, no deps,'' since it provides a
\texttt{.d.ts} declaration file and has no dependencies.
\cref{fig:cs-ieee754} shows the original declaration for the \texttt{write}
function on \cref{line:ieee754-original}, and the handwritten, ground truth signature on \cref{line:ieee754-groundtruth}.

Consider the \deeptyper{} solution: the compiler emits an error on \cref{line:ieee754-dt}, because a function is being assigned to a variable of type \texttt{void}.
However, even if that error is fixed, there is another, more subtle error not detected by the compiler: the \texttt{isLE} parameter on \cref{line:ieee754-dt-isle} is incorrectly annotated as \texttt{number}, not \texttt{boolean}.
Because this is compatible with the body of the function, there is no error.
(The \texttt{Buffer} annotation is valid, despite not matching the ground truth
\texttt{Uint8Array}, because \texttt{Buffer} is defined by the Node.js standard
library as a subtype of
\texttt{Uint8Array}.)

The \incoder{} solution on \cref{line:ieee754-ic} type checks successfully.
It also uses the \texttt{Buffer} type for \texttt{buffer}, and it uses \texttt{any} instead of \texttt{number} for the \texttt{value} parameter on \cref{line:ieee754-ic-value}.
The \texttt{any} annotation may cause run-time errors if the function is called with arguments of the wrong type.

\subsubsection{Run-Time Type Assertions}\label{sec:case-study-type-guard}

\begin{figure}[t]
\centering
\begin{lstlisting}
// LambdaNet solution
export default function (thingToPromisify: string) {#\label{line:type-guard-parameter}#
  if (typeof thingToPromisify === 'function') {#\label{line:type-guard-guard}#
    return promisify(thingToPromisify)#\label{line:type-guard-call}#
  }
  throw new TypeError('Can only promisify functions or objects')#\label{line:type-guard-exception}#
};
    \end{lstlisting}
\caption{The \lambdanet{} solution for a function adapted from the \texttt{@gar/promisify} package.}\label{fig:cs-type-guard}
\end{figure}

The \texttt{@gar/promisify} package,\footnote{\url{https://www.npmjs.com/package/@gar/promisify}} simplified and shown in \cref{fig:cs-type-guard}, is another example where a program type checks, but is incorrect.
The example exports a function that takes an argument \texttt{thingToPromisify},
type annotated as \texttt{string} by \lambdanet{}.
\cref{line:type-guard-guard} performs a run-time type check with the \texttt{typeof} operator.
This ensures that \texttt{thingToPromisify} is a function on \cref{line:type-guard-call}, which is what the \texttt{promisify} function, defined by Node.js, expects.
If \texttt{thingToPromisify} is not a function, the exception on \cref{line:type-guard-exception} is thrown.

The example type checks successfully, because the TypeScript compiler treats the \texttt{typeof} check as a type guard, and reasons that on \cref{line:type-guard-call}, the \texttt{thingToPromisify} variable has been narrowed\footnote{\url{https://www.typescriptlang.org/docs/handbook/2/narrowing.html}} to a more specific type.
However, because \texttt{thingToPromisify} is annotated as \texttt{string}, the type guard always returns false.
Therefore, \cref{line:type-guard-call} is actually unreachable, so the exception on \cref{line:type-guard-exception} is always thrown.

\subsubsection{Variable Used as Two Different Types}\label{sec:case-study-for-loops}

\begin{figure}[t]
\centering
\begin{lstlisting}
export default function(arr: any[]) {
  var len: number = arr.length;
  var o: object = {};
  var i: number;#\label{line:for-index}#

  for (i = 0; i < len; i += 1) { ... }#\label{line:for-loop}#

  for (i in o) { ... }#\label{line:for-in-loop}#
};
    \end{lstlisting}
\caption{The \lambdanet{} solution for a function adapted from the
\texttt{array-unique} package.}\label{fig:cs-for-loop}
\end{figure}

The example in \cref{fig:cs-for-loop} is adapted from the \texttt{array-unique}
package.\footnote{\url{https://www.npmjs.com/package/array-unique}}
The example contains two \texttt{for} loops: a traditional, counter-based
\texttt{for} loop on \cref{line:for-loop}, and a \texttt{for...in} loop on
\cref{line:for-in-loop} that iterates over all enumerable string properties of an object.
Both loops share the same loop variable, \texttt{i}, defined on \cref{line:for-index} and annotated as \texttt{number} by \lambdanet{}.

The use of \texttt{i} on \cref{line:for-in-loop} causes a type error, as \texttt{for...in} loops require the loop variable to be \texttt{string}.
However, changing the annotation on \cref{line:for-index} to \texttt{string}
causes a type error on \cref{line:for-loop}, as counter-based \texttt{for} loops require the loop variable to be \texttt{number}.
One solution is to use the \texttt{any} annotation, and another is to use the union type \mbox{\texttt{number | string}}.
Ultimately, the correct solution is to define separate loop variables; this example highlights that code written in JavaScript may need to be refactored for TypeScript.

\section{Discussion}\label{sec:discussion}

\subparagraph*{How should type prediction models be evaluated?}
Prior work has used accuracy to evaluate type prediction models, but in this
paper, we argue that we should instead type check the generated code. However,
we acknowledge that our proposed metric also has limitations: code may type
check with trivial annotations (\eg, \texttt{any} or \texttt{Function}) that provide little benefit
to the programmer. Furthermore, type correctness does not necessarily mean
the type annotations are correct: \texttt{any} can hide type errors that are
only encountered at run time.

We do not claim that our metric is the final word on evaluating type prediction
models, but we believe it is an improvement over accuracy. We hope this paper
can spark a discussion on how machine learning for type migration should be
evaluated.

\subparagraph*{Can slightly wrong type annotations be useful?}
A type prediction model may suggest types that are slightly wrong and easily
fixable by a programmer, but fail to type check. However, a tool that produces
hundreds or thousands of slightly wrong type annotations would overwhelm the
programmer, and we believe it is important to build tools that try to produce
fewer errors. On the other hand, slightly wrong type annotations may still
provide value, but we would need to define what ``slightly wrong'' means and how
to measure it. Without a tool like \system{}, which weaves type
annotations into code and type checks the result, it would not be possible to
ask these questions.

\subparagraph*{Should we evaluate on JavaScript or TypeScript programs?}
We choose to evaluate on JavaScript programs, so our dataset deviates from prior
work, which only considered TypeScript. Our motivating problem is not to recover
type annotations for TypeScript programs that already type check, but to migrate
untyped JavaScript programs to type-annotated TypeScript.
For this problem, type prediction on its own is not enough, and
other steps and further refactoring may be required.

Our methodology makes it possible to evaluate performance on code without known
type annotations, \ie, code that has never been typed before. In contrast, prior
work required the benchmarks to have ground truth type annotations. Our approach
also reduces the likelihood of training data leaking into the test set.

However, there may be scenarios where a type prediction model is used to
generate type annotations for a partially annotated TypeScript project. In these
situations, type migration would likely not require additional refactoring
steps.

\subparagraph*{Can we fully automate type migration?}
Our results show that automatically predicting type annotations is a challenging
task and much work remains to be done. Furthermore, migrating JavaScript to
TypeScript involves more than just adding type annotations: the two languages
are different and some refactoring may be required. The models we evaluate in
this paper do not refactor code, and we believe it is unlikely for automated
type migration to be perfect. Thus, some manual refactoring will always be
necessary for certain kinds of code, but we hope that tools can reduce the
overall burden on programmers.

\section{Related Work}\label{sec:related-work}

There are many constraint-based approaches to type migration for the gradually
typed lambda calculus (GTLC) and some modest extensions. The earliest approach
was a variation of unification-based type inference~\cite{siek:gti}, and more
recent work uses a wide range of
techniques~\cite{campora:migrating,castagna:perspective,garcia:principal-gradual-types,migeed:decidable,miyazaki:dti,phipps-costin:typewhich}.
Since these approaches are based on programming language semantics, they produce
sound results, which is their key advantage over learning-based approaches.
However, these would require significant work to scale to complex programming
languages such as JavaScript.

There are also several constraint-based approaches to type inference for larger
languages. Anderson et al.~\cite{anderson:inference} presents type inference for a small
fragment of JavaScript, but is not designed for gradual typing.
Rastogi et al.~\cite{rastogi:gti} infer gradual types for ActionScript to improve performance.
More recently, Chandra et al.~\cite{chandra:static-js} infer types for JavaScript programs
with the goal of compiling them to run efficiently on low-powered devices;
their approach is not gradual by design and deliberately rejects certain
programs. DRuby~\cite{furr:druby} infers types for Ruby and treats type
annotations in a novel way: inference assumes that annotations are correct, and
defers checking them to runtime.

Although this paper focuses on type migration for TypeScript, there are several
other gradual type systems for
JavaScript~\cite{chugh:systemd,guha:flowtypes,lerner:tejas,vekris:two-phase-typing}.
These languages do not have support for type inference and do not provide tools for
type migration. Instead, like Typed Racket~\cite{th:typed-scheme}, they require programmers to manually migrate their code
to add types. However, there are tools that
use dynamic profiles to infer types for these type
systems~\cite{an:rubydust,furr:pruby,saftoiu:jstrace}.

Even when constraint-based type inference succeeds in a gradually typed
language, it can fail to produce the kinds of types that programmers write, \eg{},
named types, instead of the most general structural type for every
annotation. Soft Scheme~\cite{cartwright:soft-typing} infers types for Scheme
programs, but Flanagan~\cite[p. 41]{flanagan:thesis} reports that it produces
unintuitive types.
For Ruby, InferDL~\cite{kazerounian:inferdl} uses hand-coded
heuristics to infer more natural types, and SimTyper~\cite{kazerounian:simtyper}
uses machine learning to predict equalities between structural types and more
natural types.

\lambdanet{}~\cite{wei:lambdanet} and \deeptyper{}~\cite{hellendoorn:dlti} are two
different approaches for predicting types for TypeScript and JavaScript programs.
This paper evaluates using both of them in its type migration pipeline.
We discuss them at length in \cref{sec:bg-deeptyper,sec:bg-lambdanet}.
NL2Type~\cite{malik:nl2type} is another system for predicting JavaScript types that
improves on DeepTyper.

\enlargethispage{-1\baselineskip}
There are also type prediction systems for Python. TypeWriter~\cite{pradel:typewriter}
is notable because it also asks if the resulting Python program type checks.
If it does not, it searches its solution space for an alternative typing.
A distinction between Python type systems and TypeScript is that Python
code is predominantly nominally typed: the type of a variable is either a builtin
type or a class, whereas TypeScript uses structural types.

DiverseTyper~\cite{jesse:diversetyper} is a recently published work that
predicts both built-in and user-defined types for TypeScript and achieves
state-of-the-art accuracy on type prediction. DiverseTyper builds on
TypeBert~\cite{jesse:typebert}, which trains a BERT-based model to predict
types. Although this paper does not evaluate these models, they are
most closely related to \incoder{}~\cite{fried:incoder}, which is a
general-purpose code generation model that we do evaluate.

\section{Conclusion}\label{sec:conclusion}

In this paper, we set out to answer the question: \emph{do deep-learning-based type annotation prediction models produce TypeScript types that type check?}
To answer this question, we build \system{}, a type migration tool that automatically converts JavaScript projects into TypeScript.
\system{} uses a type annotation prediction model, but does the work of ``weaving'' predicted types into JavaScript code.
It also automates other steps, such as converting JavaScript projects to \es{}
module notation.
Finally, \system{} runs the TypeScript compiler to type check the generated
code.
\system{} is designed so that any type prediction model can be plugged in, and we use three very different models: \deeptyper{}, \lambdanet{}, and \incoder{}.

In addition to building \system{}, we also present a dataset of \totalpkgs{} widely used JavaScript packages that are suitable for type migration.
Every package in our dataset has typed dependencies and many of them have never been typed before.
With this dataset, we evaluate \system{} with all three type prediction models.

The results are mixed.
If we ask, ``How many packages type check when migrated to TypeScript?'' we find that most packages have some type errors.
However, we also ask, ``How many files are error free?'' and the result is more promising.
We find that most files have no type errors, which means that programmers
performing type migration can focus their attention on a smaller number of files.

Our case studies highlight two insights: (1)~certain patterns in JavaScript do
not make sense in TypeScript, so a migration may require manual rewriting of the
code; and (2)~there are cases where programs successfully type check but still
have run-time errors.

We believe that currently, while type prediction cannot always reliably migrate JavaScript to TypeScript, it can still be a powerful tool.

\subparagraph*{Future Work.}
There are several directions we would like to explore in future work.
First, we would like to improve dataset quality. We observed projects
that were trivially typable: there were few declarations to annotate, or the
annotations were mostly primitive types, so those projects often type
checked successfully.
Second, we are interested in exploring different evaluation criteria for type
prediction models. We believe that type checking the output of these models is
only the first step, and that it may be necessary to evaluate the run-time
behavior of migrated programs. Additionally, there may be utility in permitting
``slightly wrong'' type annotations.
Finally, we would like to examine other deep learning models and type migration
tasks beyond type annotation prediction.

\enlargethispage{1.8\baselineskip}
\bibliography{p037-Yee}

\begin{thebibliography}{10}

\bibitem{an:rubydust}
Jong-hoon~(David) An, Avik Chaudhuri, Jeffrey~S. Foster, and Michael Hicks.
\newblock {Dynamic Inference of Static Types for Ruby}.
\newblock In {\em Principles of Programming Languages (POPL)}, 2011.
\newblock \href {https://doi.org/10.1145/1926385.1926437}
  {\path{doi:10.1145/1926385.1926437}}.

\bibitem{anderson:inference}
Christopher Anderson, Paola Giannini, and Sophia Drossopoulou.
\newblock {Towards Type Inference for JavaScript}.
\newblock In {\em European Conference on Object-Oriented Programming (ECOOP)},
  2005.
\newblock \href {https://doi.org/10.1007/11531142_19}
  {\path{doi:10.1007/11531142_19}}.

\bibitem{heap:ts}
Luke Autry.
\newblock {How we failed, then succeeded, at migrating to TypeScript}.
\newblock \url{https://heap.io/blog/migrating-to-typescript}, 2019.
\newblock Accessed: 2022-12-01.

\bibitem{bavarian:openai}
Mohammad Bavarian, Heewoo Jun, Nikolas Tezak, John Schulman, Christine
  McLeavey, Jerry Tworek, and Mark Chen.
\newblock {Efficient Training of Language Models to Fill in the Middle}, 2022.
\newblock \href {https://doi.org/10.48550/arXiv.2207.14255}
  {\path{doi:10.48550/arXiv.2207.14255}}.

\bibitem{benallal:santacoder}
Loubna {Ben Allal}, Raymond Li, Denis Kocetkov, Chenghao Mou, Christopher
  Akiki, Carlos~Munoz Ferrandis, Niklas Muennighoff, Mayank Mishra, Alex Gu,
  Manan Dey, Logesh~Kumar Umapathi, Carolyn~Jane Anderson, Yangtian Zi,
  Joel~Lamy Poirier, Hailey Schoelkopf, Sergey Troshin, Dmitry Abulkhanov,
  Manuel Romero, Michael Lappert, Francesco De~Toni, Bernardo~García del Río,
  Qian Liu, Shamik Bose, Urvashi Bhattacharyya, Terry~Yue Zhuo, Ian Yu, Paulo
  Villegas, Marco Zocca, Sourab Mangrulkar, David Lansky, Huu Nguyen, Danish
  Contractor, Luis Villa, Jia Li, Dzmitry Bahdanau, Yacine Jernite, Sean
  Hughes, Daniel Fried, Arjun Guha, Harm de~Vries, and Leandro von Werra.
\newblock {SantaCoder: don't reach for the stars!}, 2023.
\newblock \href {https://doi.org/10.48550/arXiv.2301.03988}
  {\path{doi:10.48550/arXiv.2301.03988}}.

\bibitem{bierman:ts}
Gavin Bierman, Mart{\'i}n Abadi, and Mads Torgersen.
\newblock {Understanding TypeScript}.
\newblock In {\em European Conference on Object-Oriented Programming (ECOOP)},
  2014.
\newblock \href {https://doi.org/10.1007/978-3-662-44202-9_11}
  {\path{doi:10.1007/978-3-662-44202-9_11}}.

\bibitem{bonnaire-sergeant:typed-clojure}
Ambrose Bonnaire-Sergeant, Rowan Davies, and Sam Tobin-Hochstadt.
\newblock {Practical Optional Types for Clojure}.
\newblock In {\em European Symposium on Programming (ESOP)}, 2016.
\newblock \href {https://doi.org/10.1007/978-3-662-49498-1_4}
  {\path{doi:10.1007/978-3-662-49498-1_4}}.

\bibitem{netflix:ts}
Ryan Burgess, Joe King, Stacy London, Sumana Mohan, and Jem Young.
\newblock {TypeScript migration - Strict type of cocktails}.
\newblock
  \url{https://frontendhappyhour.com/episodes/typescript-migration-strict-type-of-cocktails},
  2022.
\newblock Accessed: 2022-12-01.

\bibitem{campora:migrating}
John~Peter Campora, Sheng Chen, Martin Erwig, and Eric Walkingshaw.
\newblock {Migrating Gradual Types}.
\newblock {\em Proc. ACM Program. Lang.}, 2(POPL), 2018.
\newblock \href {https://doi.org/10.1145/3158103} {\path{doi:10.1145/3158103}}.

\bibitem{cartwright:soft-typing}
Robert Cartwright and Mike Fagan.
\newblock {Soft Typing}.
\newblock In {\em Programming Language Design and Implementation (PLDI)}, 1991.
\newblock \href {https://doi.org/10.1145/113445.113469}
  {\path{doi:10.1145/113445.113469}}.

\bibitem{cassola:gradual-elixir}
Mauricio Cassola, Agust\'{\i}n Talagorria, Alberto Pardo, and Marcos Viera.
\newblock {A Gradual Type System for Elixir}.
\newblock In {\em Brazilian Symposium on Context-Oriented Programming and
  Advanced Modularity (SBLP)}, 2020.
\newblock \href {https://doi.org/10.1145/3427081.3427084}
  {\path{doi:10.1145/3427081.3427084}}.

\bibitem{castagna:perspective}
Giuseppe Castagna, Victor Lanvin, Tommaso Petrucciani, and Jeremy~G. Siek.
\newblock {Gradual Typing: A New Perspective}.
\newblock {\em Proc. ACM Program. Lang.}, 3(POPL), 2019.
\newblock \href {https://doi.org/10.1145/3290329} {\path{doi:10.1145/3290329}}.

\bibitem{chandra:static-js}
Satish Chandra, Colin~S. Gordon, Jean-Baptiste Jeannin, Cole Schlesinger, Manu
  Sridharan, Frank Tip, and Youngil Choi.
\newblock {Type Inference for Static Compilation of JavaScript}.
\newblock In {\em Object-Oriented Programming Systems Languages and
  Applications (OOPSLA)}, 2016.
\newblock \href {https://doi.org/10.1145/2983990.2984017}
  {\path{doi:10.1145/2983990.2984017}}.

\bibitem{chauduri:fb-flow}
Avik Chaudhuri, Panagiotis Vekris, Sam Goldman, Marshall Roch, and Gabriel
  Levi.
\newblock {Fast and Precise Type Checking for JavaScript}.
\newblock {\em Proc. ACM Program. Lang.}, 1(OOPSLA), 2017.
\newblock \href {https://doi.org/10.1145/3133872} {\path{doi:10.1145/3133872}}.

\bibitem{chugh:systemd}
Ravi Chugh, David Herman, and Ranjit Jhala.
\newblock {Dependent Types for JavaScript}.
\newblock In {\em Object-Oriented Programming Systems Languages and
  Applications (OOPSLA)}, 2012.
\newblock \href {https://doi.org/10.1145/2384616.2384659}
  {\path{doi:10.1145/2384616.2384659}}.

\bibitem{feldthaus:ts-interface}
Asger Feldthaus and Anders M\o{}ller.
\newblock {Checking Correctness of TypeScript Interfaces for JavaScript
  Libraries}.
\newblock In {\em Object-Oriented Programming Systems Languages and
  Applications (OOPSLA)}, 2014.
\newblock \href {https://doi.org/10.1145/2660193.2660215}
  {\path{doi:10.1145/2660193.2660215}}.

\bibitem{flanagan:thesis}
Cormac Flanagan.
\newblock {\em {Effective Static Debugging via Componential Set-based
  Analysis}}.
\newblock PhD thesis, Rice University, 1997.
\newblock URL: \url{https://users.soe.ucsc.edu/~cormac/papers/thesis.pdf}.

\bibitem{fried:incoder}
Daniel Fried, Armen Aghajanyan, Jessy Lin, Sida Wang, Eric Wallace, Freda Shi,
  Ruiqi Zhong, Scott Yih, Luke Zettlemoyer, and Mike Lewis.
\newblock {InCoder: A Generative Model for Code Infilling and Synthesis}.
\newblock In {\em International Conference on Learning Representations (ICLR)},
  2023.
\newblock \href {https://doi.org/10.48550/arXiv.2204.05999}
  {\path{doi:10.48550/arXiv.2204.05999}}.

\bibitem{furr:pruby}
Michael Furr, Jong-hoon~(David) An, and Jeffrey~S. Foster.
\newblock {Profile-Guided Static Typing for Dynamic Scripting Languages}.
\newblock In {\em Object-Oriented Programming Systems Languages and
  Applications (OOPSLA)}, 2009.
\newblock \href {https://doi.org/10.1145/1640089.1640110}
  {\path{doi:10.1145/1640089.1640110}}.

\bibitem{furr:druby}
Michael Furr, Jong-hoon~(David) An, Jeffrey~S. Foster, and Michael Hicks.
\newblock {Static Type Inference for Ruby}.
\newblock In {\em Symposium on Applied Computing (SAC)}, 2009.
\newblock \href {https://doi.org/10.1145/1529282.1529700}
  {\path{doi:10.1145/1529282.1529700}}.

\bibitem{garcia:principal-gradual-types}
Ronald Garcia and Matteo Cimini.
\newblock {Principal Type Schemes for Gradual Programs}.
\newblock In {\em Principles of Programming Languages (POPL)}, 2015.
\newblock \href {https://doi.org/10.1145/2676726.2676992}
  {\path{doi:10.1145/2676726.2676992}}.

\bibitem{guha:flowtypes}
Arjun Guha, Claudiu Saftoiu, and Shriram Krishnamurthi.
\newblock {Typing Local Control and State Using Flow Analysis}.
\newblock In {\em European Symposium on Programming (ESOP)}, 2011.
\newblock \href {https://doi.org/10.1007/978-3-642-19718-5_14}
  {\path{doi:10.1007/978-3-642-19718-5_14}}.

\bibitem{hellendoorn:dlti}
Vincent~J. Hellendoorn, Christian Bird, Earl~T. Barr, and Miltiadis Allamanis.
\newblock {Deep Learning Type Inference}.
\newblock In {\em European Software Engineering Conference/Foundations of
  Software Engineering (ESEC/FSE)}, 2018.
\newblock \href {https://doi.org/10.1145/3236024.3236051}
  {\path{doi:10.1145/3236024.3236051}}.

\bibitem{jesse:diversetyper}
Kevin Jesse, Premkumar Devanbu, and Anand~Ashok Sawant.
\newblock {Learning To Predict User-Defined Types}.
\newblock {\em IEEE Transactions on Software Engineering (TSE)}, 2022.
\newblock \href {https://doi.org/10.1109/TSE.2022.3178945}
  {\path{doi:10.1109/TSE.2022.3178945}}.

\bibitem{jesse:typebert}
Kevin Jesse, Premkumar~T. Devanbu, and Toufique Ahmed.
\newblock {Learning Type Annotation: Is Big Data Enough?}
\newblock In {\em European Software Engineering Conference/Foundations of
  Software Engineering (ESEC/FSE)}, 2021.
\newblock \href {https://doi.org/10.1145/3468264.3473135}
  {\path{doi:10.1145/3468264.3473135}}.

\bibitem{kazerounian:simtyper}
Milod Kazerounian, Jeffrey~S. Foster, and Bonan Min.
\newblock {SimTyper: Sound Type Inference for Ruby Using Type Equality
  Prediction}.
\newblock {\em Proc. ACM Program. Lang.}, 5(OOPSLA), 2021.
\newblock \href {https://doi.org/10.1145/3485483} {\path{doi:10.1145/3485483}}.

\bibitem{kazerounian:inferdl}
Milod Kazerounian, Brianna~M. Ren, and Jeffrey~S. Foster.
\newblock {Sound, Heuristic Type Annotation Inference for Ruby}.
\newblock In {\em Dynamic Languages Symposium (DLS)}, 2020.
\newblock \href {https://doi.org/10.1145/3426422.3426985}
  {\path{doi:10.1145/3426422.3426985}}.

\bibitem{kristensen:tsd}
Erik~Krogh Kristensen and Anders M{\o}ller.
\newblock {Inference and Evolution of TypeScript Declaration Files}.
\newblock In {\em Fundamental Approaches to Software Engineering (FASE)}, 2017.
\newblock \href {https://doi.org/10.1007/978-3-662-54494-5_6}
  {\path{doi:10.1007/978-3-662-54494-5_6}}.

\bibitem{kristensen:tts}
Erik~Krogh Kristensen and Anders M\o{}ller.
\newblock {Type Test Scripts for TypeScript Testing}.
\newblock {\em Proc. ACM Program. Lang.}, 1(OOPSLA), 2017.
\newblock \href {https://doi.org/10.1145/3133914} {\path{doi:10.1145/3133914}}.

\bibitem{lerner:tejas}
Benjamin~S. Lerner, Joe~Gibbs Politz, Arjun Guha, and Shriram Krishnamurthi.
\newblock {TeJaS: Retrofitting Type Systems for JavaScript}.
\newblock In {\em Dynamic Languages Symposium (DLS)}, 2013.
\newblock \href {https://doi.org/10.1145/2578856.2508170}
  {\path{doi:10.1145/2578856.2508170}}.

\bibitem{lu:static-python}
Kuang-Chen Lu, Ben Greenman, Carl Meyer, Dino Viehland, Aniket Panse, and
  Shriram Krishnamurthi.
\newblock {Gradual Soundness: Lessons from Static Python}.
\newblock {\em The Art, Science, and Engineering of Programming}, 7(1), 2022.
\newblock \href {https://doi.org/10.22152/programming-journal.org/2023/7/2}
  {\path{doi:10.22152/programming-journal.org/2023/7/2}}.

\bibitem{malik:nl2type}
Rabee~Sohail Malik, Jibesh Patra, and Michael Pradel.
\newblock {NL2Type: Inferring JavaScript Function Types from Natural Language
  Information}.
\newblock In {\em International Conference on Software Engineering (ICSE)},
  2019.
\newblock \href {https://doi.org/10.1109/ICSE.2019.00045}
  {\path{doi:10.1109/ICSE.2019.00045}}.

\bibitem{pyre}
{Meta Platforms, Inc.}
\newblock {Pyre: A performant type-checker for Python 3}.
\newblock \url{https://pyre-check.org/}.
\newblock Accessed: 2022-12-01.

\bibitem{migeed:decidable}
Zeina Migeed and Jens Palsberg.
\newblock {What Is Decidable about Gradual Types?}
\newblock {\em Proc. ACM Program. Lang.}, 4(POPL), 2020.
\newblock \href {https://doi.org/10.1145/3371097} {\path{doi:10.1145/3371097}}.

\bibitem{miyazaki:dti}
Yusuke Miyazaki, Taro Sekiyama, and Atsushi Igarashi.
\newblock {Dynamic Type Inference for Gradual Hindley--Milner Typing}.
\newblock {\em Proc. ACM Program. Lang.}, 3(POPL), 2019.
\newblock \href {https://doi.org/10.1145/3290331} {\path{doi:10.1145/3290331}}.

\bibitem{abacus:ts}
Thomas Moore.
\newblock {How We Completed a (Partial) TypeScript Migration In Six Months}.
\newblock
  \href{https://blog.abacus.com/how-we-completed-a-partial-typescript-migration-in-six-months/}{\texttt{https://blog.abacus.com/\allowbreak{}how\allowbreak{}-we\allowbreak{}-completed\allowbreak{}-a\allowbreak{}-partial\allowbreak{}-typescript\allowbreak{}-migration\allowbreak{}-in\allowbreak{}-six\allowbreak{}-months/}},
  2019.
\newblock Accessed: 2022-12-01.

\bibitem{ottoni:hhvm}
Guilherme Ottoni.
\newblock {HHVM JIT: A Profile-Guided, Region-Based Compiler for PHP and Hack}.
\newblock In {\em Programming Language Design and Implementation (PLDI)}, 2018.
\newblock \href {https://doi.org/10.1145/3192366.3192374}
  {\path{doi:10.1145/3192366.3192374}}.

\bibitem{pandi:opttyper}
Irene~Vlassi Pandi, Earl~T. Barr, Andrew~D. Gordon, and Charles Sutton.
\newblock {OptTyper: Probabilistic Type Inference by Optimising Logical and
  Natural Constraints}, 2021.
\newblock \href {https://doi.org/10.48550/arXiv.2004.00348}
  {\path{doi:10.48550/arXiv.2004.00348}}.

\bibitem{quip:ts}
Mihai Parparita.
\newblock {The Road to TypeScript at Quip, Part Two}.
\newblock \url{https://quip.com/blog/the-road-to-typescript-at-quip-part-two},
  2020.
\newblock Accessed: 2022-12-01.

\bibitem{phipps-costin:typewhich}
Luna Phipps-Costin, Carolyn~Jane Anderson, Michael Greenberg, and Arjun Guha.
\newblock {Solver-Based Gradual Type Migration}.
\newblock {\em Proc. ACM Program. Lang.}, 5(OOPSLA), 2021.
\newblock \href {https://doi.org/10.1145/3485488} {\path{doi:10.1145/3485488}}.

\bibitem{pradel:typewriter}
Michael Pradel, Georgios Gousios, Jason Liu, and Satish Chandra.
\newblock {TypeWriter: Neural Type Prediction with Search-Based Validation}.
\newblock In {\em European Software Engineering Conference/Foundations of
  Software Engineering (ESEC/FSE)}, 2020.
\newblock \href {https://doi.org/10.1145/3368089.3409715}
  {\path{doi:10.1145/3368089.3409715}}.

\bibitem{rastogi:gti}
Aseem Rastogi, Avik Chaudhuri, and Basil Hosmer.
\newblock {The Ins and Outs of Gradual Type Inference}.
\newblock In {\em Principles of Programming Languages (POPL)}, 2012.
\newblock \href {https://doi.org/10.1145/2103656.2103714}
  {\path{doi:10.1145/2103656.2103714}}.

\bibitem{slack:ts}
Felix Rieseberg.
\newblock {TypeScript at Slack}.
\newblock \url{https://slack.engineering/typescript-at-slack/}, 2017.
\newblock Accessed: 2022-12-01.

\bibitem{airbnb:ts-migrate}
Sergii Rudenko.
\newblock {ts-migrate: A Tool for Migrating to TypeScript at Scale}.
\newblock
  \href{https://medium.com/airbnb-engineering/ts-migrate-a-tool-for-migrating-to-typescript-at-scale-cd23bfeb5cc}{\texttt{https://medium.com/airbnb\allowbreak{}-engineering/\allowbreak{}ts\allowbreak{}-migrate\allowbreak{}-a\allowbreak{}-tool\allowbreak{}-for\allowbreak{}-migrating\allowbreak{}-to\allowbreak{}-type\allowbreak{}script\allowbreak{}-at\allowbreak{}-scale\allowbreak{}-cd23bfeb5cc}},
  2020.
\newblock Accessed: 2022-12-01.

\bibitem{saftoiu:jstrace}
Claudiu Saftoiu.
\newblock {JSTrace: Run-time Type Discovery for JavaScript}.
\newblock Master's thesis, Brown University, 2010.
\newblock URL:
  \url{https://cs.brown.edu/research/pubs/theses/ugrad/2010/saftoiu.pdf}.

\bibitem{siek:gtlc}
Jeremy~G. Siek and Walid Taha.
\newblock {Gradual Typing for Functional Languages}.
\newblock In {\em Scheme and Functional Programming Workshop}, 2006.
\newblock URL: \url{http://schemeworkshop.org/2006/13-siek.pdf}.

\bibitem{siek:gti}
Jeremy~G. Siek and Manish Vachharajani.
\newblock {Gradual Typing with Unification-Based Inference}.
\newblock In {\em Dynamic Languages Symposium (DLS)}, 2008.
\newblock \href {https://doi.org/10.1145/1408681.1408688}
  {\path{doi:10.1145/1408681.1408688}}.

\bibitem{th:typed-scheme}
Sam Tobin-Hochstadt and Matthias Felleisen.
\newblock {The Design and Implementation of Typed Scheme}.
\newblock In {\em Principles of Programming Languages (POPL)}, 2008.
\newblock \href {https://doi.org/10.1145/1328438.1328486}
  {\path{doi:10.1145/1328438.1328486}}.

\bibitem{vekris:two-phase-typing}
Panagiotis Vekris, Benjamin Cosman, and Ranjit Jhala.
\newblock {Trust, but Verify: Two-Phase Typing for Dynamic Languages}.
\newblock In {\em European Conference on Object-Oriented Programming (ECOOP)},
  2015.
\newblock \href {https://doi.org/10.4230/LIPIcs.ECOOP.2015.52}
  {\path{doi:10.4230/LIPIcs.ECOOP.2015.52}}.

\bibitem{wei:lambdanet}
Jiayi Wei, Maruth Goyal, Greg Durrett, and Isil Dillig.
\newblock {LambdaNet: Probabilistic Type Inference using Graph Neural
  Networks}.
\newblock In {\em International Conference on Learning Representations (ICLR)},
  2020.
\newblock \href {https://doi.org/10.48550/arXiv.2005.02161}
  {\path{doi:10.48550/arXiv.2005.02161}}.

\bibitem{williams:ts-tpd}
Jack Williams, J.~Garrett Morris, Philip Wadler, and Jakub Zalewski.
\newblock {Mixed Messages: Measuring Conformance and Non-Interference in
  TypeScript}.
\newblock In {\em European Conference on Object-Oriented Programming (ECOOP)},
  2017.
\newblock \href {https://doi.org/10.4230/LIPIcs.ECOOP.2017.28}
  {\path{doi:10.4230/LIPIcs.ECOOP.2017.28}}.

\bibitem{stripe:sorbet}
Jake Zimmerman.
\newblock {Sorbet: Stripe's type checker for Ruby}.
\newblock \url{https://stripe.com/blog/sorbet-stripes-type-checker-for-ruby},
  2022.
\newblock Accessed: 2022-12-01.

\end{thebibliography}

\end{document}